\documentclass[12pt]{article}

\usepackage{amsmath,amsfonts,enumerate,array}
\usepackage{graphicx}
\usepackage[usenames]{color}
\usepackage[english]{babel}
\usepackage{fancybox}
\usepackage{chngpage} % allows for temporary adjustment of side margins

\newcommand{\captionfonts}{\small}

\makeatletter  % Allow the use of @ in command names
\long\def\@makecaption#1#2{%
  \vskip\abovecaptionskip
  \sbox\@tempboxa{{\captionfonts #1: #2}}%
 \ifdim \wd\@tempboxa >\hsize
    {\captionfonts #1: #2\par}
  \else
    \hbox to\hsize{\hfil\box\@tempboxa\hfil}%
  \fi
  \vskip\belowcaptionskip}
\makeatother   % Cancel the effect of \makeatletter

\usepackage{mciteplus}

\usepackage[colorlinks=true,      linkcolor=blue,      urlcolor=blue,      filecolor=blue,      citecolor=blue,
      pdfstartview=FitH,     pdfpagemode=UseNone,      bookmarksopen=true]{hyperref}  % LINKS
      
\usepackage[all]{hypcap}  % LINKS
      
\title{Full action of two deformation operators in the D1D5 CFT}

\begin{document}

\numberwithin{equation}{section}

%%%%%%%%%%%%%%%%%%%%%%%%%%%%%%%%%%%%%%%%%%%%%%%%%%%%%%%%%%%%
%                       DEFINITIONS

\mathchardef\mhyphen="2D

%%%%%%%%%%%%%%%%%%%%%%%%%%%%%%%%%%%%%%%%%%%%%%%%%%%%%%%%%%%%
%                        Commands

%			Wick Contractions
\makeatletter
\newcommand{\contraction}[5][1ex]{%
  \mathchoice
    {\contraction@\displaystyle{#2}{#3}{#4}{#5}{#1}}%
    {\contraction@\textstyle{#2}{#3}{#4}{#5}{#1}}%
    {\contraction@\scriptstyle{#2}{#3}{#4}{#5}{#1}}%
    {\contraction@\scriptscriptstyle{#2}{#3}{#4}{#5}{#1}}}%
\newcommand{\contraction@}[6]{%
  \setbox0=\hbox{$#1#2$}%
  \setbox2=\hbox{$#1#3$}%
  \setbox4=\hbox{$#1#4$}%
  \setbox6=\hbox{$#1#5$}%
  \dimen0=\wd2%
  \advance\dimen0 by \wd6%
  \divide\dimen0 by 2%
  \advance\dimen0 by \wd4%
  \vbox{%
    \hbox to 0pt{%
      \kern \wd0%
      \kern 0.5\wd2%
      \contraction@@{\dimen0}{#6}%
      \hss}%
    \vskip 0.2ex%
    \vskip\ht2}}
\newcommand{\contracted}[5][1ex]{%
  \contraction[#1]{#2}{#3}{#4}{#5}\ensuremath{#2#3#4#5}}
\newcommand{\contraction@@}[3][0.06em]{%
  \hbox{%
    \vrule width #1 height 0pt depth #3%
    \vrule width #2 height 0pt depth #1%
    \vrule width #1 height 0pt depth #3%
    \relax}}
\makeatother

%			Environments
\newcommand{\be}{\begin{equation}} % Use eqnarray instead
\newcommand{\ee}{\end{equation}} % Use eqnarray insted
\newcommand{\bea}{\begin{eqnarray}\displaystyle}
\newcommand{\eea}{\end{eqnarray}}
\newcommand{\bt}{\begin{tabular}}
\newcommand{\et}{\end{tabular}}
\newcommand{\bs}{\begin{split}}
\newcommand{\es}{\end{split}}

%			Vacuum States (In Math Environments)
\newcommand{\nsnsket}{|0_{NS}\rangle^{(1)}\newotimes |0_{NS}\rangle^{(2)}}					% Double NS ket
\newcommand{\nsnsbra}{{}^{(1)}\langle 0_{NS}| \newotimes {}^{(2)}\langle 0_{NS}|}		% Double NS bra
\newcommand{\nsket}{|0_{NS}\rangle}																									% Single NS ket
\newcommand{\nsbra}{\langle 0_{NS}|}																								% Single NS bra
\newcommand{\nstket}{|0_{NS}\rangle_t}																										% t-plane NS ket
\newcommand{\nstbra}{{}_t\langle 0_{NS}|}																								% t-plane NS bra
\newcommand{\rmmket}{|0_R^-\rangle^{(1)}\newotimes |0_R^-\rangle^{(2)}}							% Double R- ket
\newcommand{\rmmbra}{{}^{(1)}\langle 0_{R,-}| \newotimes {}^{(2)}\langle 0_{R,-}|}	% Double R- bra
\newcommand{\rmket}{|0_R^-\rangle}																									% Single R- ket
\newcommand{\rmbra}{\langle 0_{R,-}|}																								% Single R- bra
\newcommand{\rmtket}{|0_R^-\rangle_t}																								% t-plane R- ket
\newcommand{\rmtbra}{{}_t\langle 0_{R,-}|}																					% t-plane R- bra
\newcommand{\rppket}{|0_R^+\rangle^{(1)}\newotimes |0_R^+\rangle^{(2)}}							% Double R+ ket
\newcommand{\rppbra}{{}^{(1)}\langle 0_{R,+}| \newotimes {}^{(2)}\langle 0_{R,+}|}	% Double R+ bra
\newcommand{\rpket}{|0_R^+\rangle}																									% Single R+ ket
\newcommand{\rpbra}{\langle 0_{R,+}|}																								% Single R+ bra
\newcommand{\rptket}{|0_R^+\rangle_t}																								% t-plane R+ ket
\newcommand{\rptbra}{{}_t\langle 0_{R,+}|}																					% t-plane R+ bra
\newcommand{\rpmket}{| 0_R^+\rangle^{(1)} \newotimes | 0_R^-\rangle^{(2)}}					% R+ on 1, R- on 2, ket
\newcommand{\rpmbra}{{}^{(1)}\langle 0_{R,+}| \newotimes {}^{(2)}\langle 0_{R,-}|}	% R+ on 1, R- on 2, bra
\newcommand{\rmpket}{| 0_R^-\rangle^{(1)} \newotimes | 0_R^+\rangle^{(2)}}					% R- on 1, R+ on 2, ket
\newcommand{\rmpbra}{{}^{(1)}\langle 0_{R,-}| \newotimes {}^{(2)}\langle 0_{R,+}|}	% R- on 1, R+ on 2, bra

% Command Shortcuts - Vacuum States %
\newcommand{\nsutvket}{|0_{NS}\rangle^{(1)}\otimes |0_{NS}\rangle^{(2)}}
\newcommand{\nsutvbra}{{}^{(1)}\langle 0_{NS}| \otimes {}^{(2)}\langle 0_{NS}|}
\newcommand{\nstvket}{|0_{NS}\rangle}
\newcommand{\nstvbra}{\langle 0_{NS}|}
\newcommand{\nstpket}{|0\rangle_t}
\newcommand{\nstpbra}{{}_t\langle 0|}
\newcommand{\rmutvket}{|0_R^-\rangle^{(1)}\otimes |0_R^-\rangle^{(2)}}
\newcommand{\rmutvbra}{{}^{(1)}\langle 0_{R,-}| \otimes {}^{(2)}\langle 0_{R,-}|} 
\newcommand{\rmtvket}{|0_R^-\rangle}
\newcommand{\rmtvbra}{\langle 0_{R,-}|}
\newcommand{\rmtpket}{|0_R^-\rangle_t}
\newcommand{\rmtpbra}{{}_t\langle 0_{R,-}|}
\newcommand{\rputvket}{|0_R^+\rangle^{(1)}\otimes |0_R^+\rangle^{(2)}}
\newcommand{\rputvbra}{{}^{(1)}\langle 0_{R,+}| \otimes {}^{(2)}\langle 0_{R,+}|} 
\newcommand{\rptvket}{|0_R^+\rangle}
\newcommand{\rptvbra}{\langle 0_{R,+}|}
\newcommand{\rptpket}{|0_R^+\rangle_t}
\newcommand{\rptpbra}{{}_t\langle 0_{R,+}|}
\newcommand{\stp}{\sigma_2^+}
\newcommand{\stm}{\sigma_2^-}

% Greek Letters %
\renewcommand{\a}{\alpha}	% For spectral flow and boson modes
\renewcommand{\b}{\beta}
\newcommand{\g}{\gamma}		% For exponential coefficients
\newcommand{\G}{\Gamma}
\renewcommand{\d}{\delta}
\newcommand{\D}{\Delta}
\renewcommand{\c}{\chi}			% For final state
\newcommand{\C}{\Chi}
\renewcommand{\P}{\Psi}
\newcommand{\s}{\sigma}		% For twist operator and cylinder spatial coordinate
\renewcommand{\S}{\Sigma}
\renewcommand{\t}{\tau}		% For cylinder temporal coordinate
\newcommand{\e}{\epsilon}
\newcommand{\n}{\nu}
\newcommand{\m}{\mu}
\renewcommand{\r}{\rho}
\renewcommand{\l}{\lambda}
\newcommand{\sh}{\,\hat\sigma\,} % For sigma-hat with good spacing

%			Math
\newcommand{\nn}{\nonumber\\} 		% New line without numbering current line
\newcommand{\newotimes}{}  				% Change this to put tensor products back in.
\newcommand{\diff}{\,\text{d}}		% For differentials
\newcommand{\h}{{1\over2}}				% Shortcut for 1/2
\newcommand{\Gf}[1]{\G \Big{(} #1 \Big{)}}	% Gamma Function Shortcut
\newcommand{\floor}[1]{\left\lfloor #1 \right\rfloor}
\newcommand{\ceil}[1]{\left\lceil #1 \right\rceil}
\newcommand{\com}[2]{[#1,\,#2]}

%      Calligraphic Font
\def\cA{{\cal A}} \def\cB{{\cal B}} \def\cC{{\cal C}}
\def\cD{{\cal D}} \def\cE{{\cal E}} \def\cF{{\cal F}}
\def\cG{{\cal G}} \def\cH{{\cal H}} \def\cI{{\cal I}}
\def\cJ{{\cal J}} \def\cK{{\cal K}} \def\cL{{\cal L}}
\def\cM{{\cal M}} \def\cN{{\cal N}} \def\cO{{\cal O}}
\def\cP{{\cal P}} \def\cQ{{\cal Q}} \def\cR{{\cal R}}
\def\cS{{\cal S}} \def\cT{{\cal T}} \def\cU{{\cal U}}
\def\cV{{\cal V}} \def\cW{{\cal W}} \def\cX{{\cal X}}
\def\cY{{\cal Y}} \def\cZ{{\cal Z}}

%				Math Bold Face
\def\mC{\mathbb{C}} \def\mP{\mathbb{P}}  
\def\mR{\mathbb{R}} \def\mZ{\mathbb{Z}} 
\def\mT{\mathbb{T}} \def\mN{\mathbb{N}}
\def\mH{\mathbb{H}} \def\mX{\mathbb{X}}
\def\CP{\mathbb{CP}}
\def\RP{\mathbb{RP}}
\def\Z{\mathbb{Z}}
\def\N{\mathbb{N}}
\def\H{\mathbb{H}}

%			Quantum Mechanics Notation
\newcommand{\Zd}{\ensuremath{ Z^{\dagger}}}
\newcommand{\Xd}{\ensuremath{ X^{\dagger}}}
\newcommand{\Ad}{\ensuremath{ A^{\dagger}}}
\newcommand{\Bd}{\ensuremath{ B^{\dagger}}}
\newcommand{\Ud}{\ensuremath{ U^{\dagger}}}
\newcommand{\Td}{\ensuremath{ T^{\dagger}}}
\newcommand{\T}[3]{\ensuremath{ #1{}^{#2}_{\phantom{#2} \! #3}}}		%general tensor with upper indices first 
\newcommand{\tr}{\operatorname{tr}}
\newcommand{\sech}{\operatorname{sech}}
\newcommand{\Spin}{\operatorname{Spin}}
\newcommand{\Sym}{\operatorname{Sym}}
\newcommand{\Com}{\operatorname{Com}}
\def\adj{\textrm{adj}}
\def\id{\textrm{id}}
\def\pb{\ov\psi}
\def\pt{\widetilde{\psi}}
\def\at{\widetilde{\a}}
\def\cb{\ov\chi}
\def\db{\bar\partial}
\def\delb{\bar\partial}
\def\dbar{\ov\partial}
\def\dag{\dagger}
\def\dalpha{{\dot\alpha}}
\def\dbeta{{\dot\beta}}
\def\dgamma{{\dot\gamma}}
\def\ddelta{{\dot\delta}}
\def\ad{{\dot\alpha}}
\def\bd{{\dot\beta}}
\def\dg{{\dot\gamma}}
\def\dd{{\dot\delta}}
\def\th{\theta}
\def\Th{\Theta}
\def\eb{{\ov \epsilon}}
\def\gb{{\ov \gamma}}
\def\wb{{\ov w}}
\def\Wb{{\ov W}}
\def\D{\Delta}
\def\DD{\Delta^\dag}
\def\Db{\ov D}
\def\ov{\overline}
\def\Slash{\, / \! \! \! \!}
\def\dslash{\partial\!\!\!/} 
\def\Dslash{D\!\!\!\!/\,\,}
\def\fslash#1{\slash\!\!\!#1}
\def\Fslash#1{\slash\!\!\!\!#1}
\def\del{\partial}
\def\delb{\bar\partial}
\newcommand{\ex}[1]{{\rm e}^{#1}} 
\def\ii{{i}}
\newcommand{\vs}[1]{\vspace{#1 mm}}
\newcommand{\ve}{{\vec{\e}}}
\newcommand{\shalf}{\frac{1}{2}}
\newcommand{\lb}{\rangle}
\newcommand{\al}{\ensuremath{\alpha'}}
\newcommand{\ap}{\ensuremath{\alpha'}}
\newcommand{\ft}[2]{{\textstyle {\frac{#1}{#2}} }}

%				Other
\newcommand{\rmd}{\mathrm{d}}
\newcommand{\rmx}{\mathrm{x}}
\def\tA{ {\widetilde A} } 
\def\one{{\hbox{\kern+.5mm 1\kern-.8mm l}}}
\def\zero{{\hbox{0\kern-1.5mm 0}}}
\def\eq#1{(\ref{#1})}
\newcommand{\secn}[1]{Section~\ref{#1}}
\newcommand{\tbl}[1]{Table~\ref{#1}}
\newcommand{\fig}{Fig.~\ref}
\def\sqi{{1\over \sqrt{2}}}
\newcommand{\hsp}{\hspace{0.5cm}}
\def\half{{\textstyle{1\over2}}}
\let\ci=\cite \let\re=\ref
\let\se=\section \let\sse=\subsection \let\ssse=\subsubsection
\newcommand{\dpb}{D$p$-brane}
\newcommand{\dpbs}{D$p$-branes}
\def\gh{{\rm gh}}
\def\sgh{{\rm sgh}}
\def\NS{{\rm NS}}
\def\R{{\rm R}}
\def\Qp{Q_{\rm P}}
\def\QP{Q_{\rm P}}
\newcommand\dott[2]{#1 \! \cdot \! #2}
\def\eo{\overline{e}}
\newcommand{\bb}{\bigskip}
\newcommand{\ac}[2]{\ensuremath{\{ #1, #2 \}}}
\renewcommand{\ell}{l}
\newcommand{\z}{\ell}
\newcommand{\bm}{\bibitem}
\newcommand{\pa}{\partial}
\newcommand{\p}{\partial}

\vspace{16mm}

 \begin{center}
{\LARGE Full action of two deformation operators\\ in the D1D5 CFT}

\vspace{18mm}
{\bf  Zaq Carson\footnote{zcarson@physics.utoronto.ca}, Shaun Hampton\footnote{hampton.197@osu.edu}, and Samir D. Mathur\footnote{mathur.16@osu.edu}
\\}
\vspace{15mm}
${}^{1}$Department of Physics,\\ University of Toronto,\\ Toronto,
Ontario, M5S 1A7, Canada\\ \vspace{8mm}
${}^{2,3}$Department of Physics,\\ The Ohio State University,\\ Columbus,
OH 43210, USA\\ 

\vspace{8mm}
\end{center}

\vspace{4mm}

\thispagestyle{empty}
\begin{abstract}

\vspace{3mm}

We are interested in thermalization in the D1D5 CFT, since this process is expected to be dual to black hole formation. We expect that the lowest order process where thermalization occurs will be at second order in the perturbation that moves us away from the orbifold point. The operator governing the deformation off of the orbifold point consists of a twist operator combined with a supercharge operator acting on this twist.  In a previous paper we computed the action of two twist operators on an arbitrary state of the CFT. In the present work we compute the action of the supercharges on these twist operators, thereby obtaining the full action of two deformation operators on an arbitrary state of the CFT. We show that the full amplitude can be related to the amplitude with just the twists through an action of the supercharge operators on the initial and final states. The essential part of this computation consists of moving the contours from the twist operators to the initial and final states; to do this one must first map the amplitude to a covering space where the twists are removed, and then map back to the original space on which the CFT is defined.

\end{abstract}
\newpage

\section{Introduction}\label{Intro}
Black holes are systems in which the dominant force is gravity, yet their evaporation is fundamentally quantum. These are two phenomena which have proven difficult to combine into a single framework and  has proven a main goal of theoretical physics. This provides us with a compelling testing ground for any attempt we make at understanding quantum gravity.  In the case of string theory, the gravitational description can be studied by investigating its CFT dual \cite{adscft}.  This dual CFT is the focus of our investigations.

While the exact dual CFT is strongly coupled, an examination of its `free' or `orbifold' point has produced many results \cite{sv,lm1, lm2,orbifold1,orbifold2,deformation,orbifold3}.  At this coupling the dual CFT consists of several symmetrized copies of a free CFT whose target space is a 1+1 dimensional sigma model.  This orbifold model has successfully reproduced the entropy and greybody factors of near-extremal black holes  \cite{radiation1}, but it cannot provide a description of their formation.  This is because the black hole formation is dual to a thermalization of the dual CFT, which does not in general occur for excitations in a free theory.

In light of this, it is advantageous to explore a marginal deformation of the CFT away from its orbifold point.  This deformation is given by the operator \cite{acm1}:
\bea
\hat{O}_{\dot A\dot B}\left(w_0,\bar{w}_0\right)=\left[{1\over2\pi i}\oint_{w_0}G^-_{\dot A}(w)\diff w\right]\left[{1\over2\pi i}\oint_{w_0}\bar{G}^-_{\dot B}(\bar w)\diff \bar w\right]\s_2^{++}\left(w_0,\bar{w}_0\right).\label{FullDeformationOperator}
\eea
The index notations are detailed in appendix \ref{ap:CFT-notation} but here we discuss the general structure of this operator.  This operator contains two key components.  The first is the twist, $\s_2$, which joins two copies of the free CFT.  If these copies were built on circles of length $2\pi R$, the twist merges them into a single CFT on a circle of length $4\pi R$. The second ingredient is the supercharge operator $G$, which is applied in both left-moving and right-moving sectors. The combination of these two gives an exactly marginal operator.

For the case when we have a single deformation operator, it was shown in  \cite{acm1,acm2,chmt1} that the supercharge contour can be removed from the twist by stretching it away until it acts on the initial and final states of the process.  This allows us to separate out the action of the `bare twist' $\s_2$ from the action of the supercharge.  However, no clear mechanism for thermalization was found at first  order in the deformation.

\begin{figure}[tbh]
\begin{center}
\includegraphics[width=0.2\columnwidth]{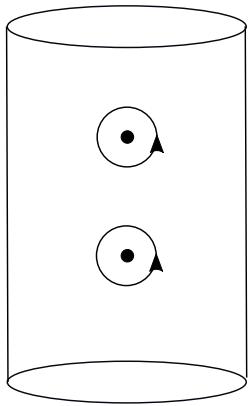}
\end{center}
\caption{The cylinder with twist insertions at $w_1$ and $w_2$ and supercharge contours circling the twist insertions.}
\label{figure1}
\end{figure}

In \cite{chm1} it was noted that we do expect the essential thermalization vertex to emerge at second order in the deformation. In \cite{chm1} the effect of two `bare twists' was computed on initial states which were either the vacuum or contained one oscillator excitation. In the present paper, we wish to include the action of the supercharges that act on these twists to make them into full deformation operators. The supercharges acting on the twist operators are depicted in Figure \ref{figure1}.

The essential process for handling the supercharges is the same as in the case where we had just one deformation operator. The result, however, is more complicated. With the case of one twist, we can unwrap the $G$ contour from the twist and change it to contours that act above and below this twist; thus we get modes of $G$ acting on the initial and final states of the amplitude. With two twists, we can again try to unwrap contours of $G$ from the twists. But this process yields terms where the contours of $G$ are stuck between the two twist operators; i.e., we get contours that do not give $G$ modes acting on the initial and final states of the amplitude.

By going to the covering space of the space where the CFT is defined, we can undo the action of the twists, and move the $G$ contours in such a way that they do act only on the initial and final states. This allows us to relate the full action of two deformation operators (i.e. with the supercharge actions included) to the amplitude with just the two bare twists. The resulting expression is the main result of this paper. We find terms where both supercharge contours act on the final state, terms where they both act on the initial state, and terms where one contour acts on the final state and one contour acts on the inital state.

Let us begin by recalling the computation \cite{chm1} that inserts two `bare twists';. i.e., the deformation operators without the supercharges.  
The case we examined was the following. One twist joins two singly-wound copies of the orbifold CFT to a single copy of the CFT living on a double circle. This is followed  by a second twist which returns the double circle back to two singly-wound copies. It was shown that when the initial CFT copies are both in a vacuum state, the result is a squeezed state of the schematic form:
\bea
\s_2^+(w_2)\s_2^+(w_1)|0\rangle &=& e^{\g^B_{mn}\a_{-m}\a_{-n}+\g^F_{rs}d_{-r}d_{-s}}|0\rangle ~\equiv~ |\chi\rangle,
\label{chi}
\eea
where the mode indices are summed over all creation operators.  The $\a$ modes are bosonic, while the $d$ modes are fermionic.  The coefficients $\g^B$ and $\g^F$ were expressed in terms of finite sums and their behavior for large indices was analyzed.

\bigskip

\bigskip

\bigskip

In \cite{chm2} we extended our analysis to include initial excitations in the 1-loop process. Our results showed that when starting with a single initial excitation on one of the copies, a final state containing a weighted linear combination of excitations acting on the squeezed state was produced. This weight is captured by the $f$ function, or transition amplitude. Schematically, our results were of the form:
\bea
\s_2^+(w_2)\s_2^+(w_1)\a_{-n}|0\rangle &=& \sum_{p}f^B_{np}\a_{-p}|\chi\rangle\nn
\s_2^+(w_2)\s_2^+(w_1)d_{-n}|0\rangle &=& \sum_{p}f^F_{np}d_{-p}|\chi\rangle,
\eea
where again our sum encompasses all creation modes.  One readily notes that this result was schematically identical to the first-order case found in \cite{acm2}.  We expect this form at all orders for the same reason we expect the bogoliubov form of $|\chi\rangle$ at all orders: Each mode on the cylinder maps to a linear combination of single modes with the same SU(2) indices in the twist-free covering space.  Since the vacuum gives a bogoliubov,  single initial excitation will give a single excitation above the bogoliubov state.

The plan of this paper is as follows.  In section \ref{CFT}, we introduce the orbifold CFT.  In section \ref{G contour}, we write the second order action of the $G$ contours on the two twist operators, $\s_2^+\s_2^+$, on the base space. We then map to the covering space which will allow us to write the $G$ contours originally circling the two twist insertions, at initial and final states. As a result of the manipulations in section \ref{G contour}, we will have two main integrals to compute labeled $\mathcal{I}_1$ and $\mathcal{I}_2$ either containing one or two $G$ contours surrounding a twist. In section \ref{computing I1}, we compute $\mathcal{I}_1$ by mapping to the cover space and mapping back. In section \ref{computing I2}, we compute $\mathcal{I}_2$ by mapping to the cover space and mapping back. In section \ref{cylinder result}, we write down the full expression of the two deformation operator as contours on the cylinder using the results of sections \ref{computing I1} and \ref{computing I2}. The expressions in section \ref{cylinder result} will contain the cover space coordinate which depends nontrivially on the cylinder coordinate. In section \ref{map inversion}, we write the covering space coordinate coming in the contour integrals in section \ref{cylinder result} as expansions in terms of the cylinder coordinate. In section \ref{cylinder modes}, we use the expansions computed in section \ref{map inversion}, to write the cylinder contours in terms of cylinder modes. In section \ref{final result}, we use the results computed in section \ref{cylinder modes}, to write down the final result of the two deformation operators in terms of modes defined on the cylinder.

\section{The orbifold CFT}\label{CFT}
Consider type IIB string theory, compactified as:
\bea
M_{9,1} &\to& M_{4,1}\times S^1\times T^4.
\eea
We then wrap $N_1$ D1 branes on $S^1$ and $N_5$ D5 branes on $S^1 \times T^4$.  We take $S^1$ to be large compared to $T^4$, so that the low energies are dominated by excitations only in the direction $S^1$.  This low-energy limit gives a $1+1$ dimensional CFT living on $S^1$.

At this point, variations in the moduli of string theory move us through the moduli space of the CFT on $S^1$.  It is conjectured that we can move to an 'orbifold point' where this CFT is free and can be described by a particularly simple sigma model \cite{orbifold2}.  We will begin in the Euclidean theory at this orbifold point.  The base space is a cylinder spanned by the coordinates $\t,\s$:
\bea
0\leq\s<2\pi,\qquad -\infty<\t<\infty.
\eea
The target space of this CFT is the symmetrized product of $N_1 N_5$ copies of $T^4$:
\bea
(T^4)^{N_1 N_5}/S_{N_1 N_5}.
\eea
Each copy gives 4 bosonic excitations and 4 fermionic excitations.  With an index $i$ ranging from 1 to 4, we label the bosonic excitations $X^i$, the left-moving fermionic excitations $\psi^i$, and the right-moving fermionic $\bar{\psi}^i$.  The total central charge is then $6 N_1 N_5$.

Fortunately, the twist operator fully factorizes into separate left-moving (holomorphic) and right-moving (antiholomorphic) sectors.  We thus constrain our analysis to the left-moving portion of (\ref{FullDeformationOperator}) and to holomorphic excitations.  The right-moving sector is completely analogous.

\subsection{NS and R vacuua}
At the orbifold point, each separate CFT copy has central charge $c=6$.  The lowest energy state in the left-moving sector for such a copy is the NS vacuum:
\bea
\nsket, &&\qquad h=0,\quad m=0,
\eea
where $h$ is the $L_0$ eigenvalue.  However, our interest lies mostly in the R sector of the CFT.  The vaccua of this sector are given by the following:
\bea
|0_R^{\pm}\rangle, &&\qquad h={1\over4}, \quad m = \pm \h\nn
|0_R\rangle,|\tilde 0_R\rangle, &&\qquad h={1\over4}, \quad m=0.\label{RamondVacuua}
\eea
The positive and neutral Ramond vacuua are defined with respect to fermion zero modes acting on the negative Ramond vacuum as follows:
\bea
\rpket & = & d^{++}_0d^{+-}_0\rmket\cr
|0_R\rangle & = & d_0^{+-}\rmket\cr
|\tilde{0}_R\rangle & = & d_0^{++}\rmket\nn
\eea
One can also relate the R and NS sectors via spectral flow \cite{spectralref}. Under spectral flow by an amount $\a$, the dimension and charge change in the following way:
\bea
h' = h +\a j + {c\a^2\over 24}\cr
j' = j +{\a c \over 12}  
\eea  
Spectral flow by a single unit in the left-moving sector produces the transformations:
\bea
\a=1: && \qquad \rmket \to \nsket, \quad \nsket \to \rpket\nn
\a=-1: && \qquad \rpket \to \nsket, \quad \nsket \to \rmket.
\eea
The other R vacuua can flow to the NS sector by first relating them to $|0_R^{\pm}\rangle$ via fermion zero modes.

\section{The amplitude and its map to the covering space}\label{G contour}

We begin by describing the cylinder coordinate \textit{w}. We then introduce the initial and final states as well as the twist operators. We record their location on the cylinder. 

Next, we note the amplitude that we have to compute. We then map this amplitude to the covering space of the CFT, where the action of the twists will be removed. 

Furthermore, since the CFT has a left (holomorphic) sector and a right (antiholomorphic) sector, our computation is completely factorized between these sectors. Therefore, we just concentrate on just the left sector; the right sector is entirely analogous.

\subsection{The Cylinder}

Now let us start with defining our cylinder coordinate
\bea
w=\t + i\s
\eea
where $\t$ is euclidean time and $\s$ is the spatial coordinate of our cylinder. The cylinder radius R has been included both coordinates so that $\s$ just becomes the angle. In our initial state at $\t=-\infty$ we have two singly wound copies of the CFT both in the negative Ramond vacuum which can be defined via spin fields applied to $NS$ vacuua for each copy:
\bea
\rmmket = S^{(1)-}(\t = -\infty)S^{(2)-}(\t = -\infty)\nsnsket\label{initial}
\eea
Each twist operator also contains a spin field:
\bea
\s^+_2(w_1)&=&S^+(w_1)\s_2(w_1)\cr
\s^+_2(w_2)&=&S^+(w_2)\s_2(w_2)
\eea
where we've taken $|w_2|>|w_1|$. At $\t \to \infty$ we have two singly wound copies of the CFT described by the state $|\chi(w_1,w_2)\rangle$ which was computed in \cite{chm1} and given schematically in (\ref{chi}).
The locations of the initial and final copies of the CFT as well as the twist insertions are tabulated below:
\bea
w&=&-\infty + i\s,~~~~ \text{Copy 1 and 2 initial}\cr
w&=&\infty + i\s,~~~~ \text{Copy 1 and 2 final}\cr
w_1&=&\t_{1} + i\s_1,~~~~ \text{First twist insertion}\cr
w_2&=&\t_{2} + i\s_2,~~~~ \text{Second twist insertion}, \quad \t_2>\t_1
\eea
Next we describe the supercharge action on the cylinder.

\subsection{Supercharge Contours on the Cylinder}
Now we describe the supercharge action on the cylinder. We note that the two copies of the deformation operator will be taken to have $SU(2)_2$ indices $\dot A, \dot B$. We will write the supercharge action at each twist as a contour surrounding that twist. Our goal is to remove the supercharge contours from these twists, and to move them to the initial and final states. This will relate the full amplitude for two deformation operators to an amplitude with just two bare twists.

The two deformation operators on the cylinder give the operator
\bea
\hat{O}_{\dot{B}}\hat{O}_{\dot{A}}&=&{1\over 2 \pi i}\oint_{w_2} dw G^-_{\dot{B}}\s_2^+(w_2){1\over 2 \pi i}\oint_{w_1} dw'G^-_{\dot{A}}(w')\s_2^+(w_1) 
\eea

Let us first stretch the contour for the operator $G^-_{\dot B}$ away from the twist on which it is applied. We will get three different contributions corresponding to three different locations:
\begin{itemize}
\item $\dot{B}$ positive direction contour for both copies at $\t > \t_2$\\
\item $\dot{B}$ negative direction contour for both copies at $\t < \t_1$\\
\item $\dot{B}$ negative direction contour outside of $\dot{A}$ contour at $\t_1$
\end{itemize}

Stretching these contours therefore gives on the cylinder:
\bea
\hat{O}_{\dot{B}}\hat{O}_{\dot{A}} &=& {1\over 2 \pi i}\int_{\s =0,\t>\t_2}^{\s=2\pi}\diff w G^-_{\dot{B}}(w)\s_2^+(w_2){1\over 2 \pi i}\oint_{w_1} \diff w'G^-_{\dot{A}}(w')\s_2^+(w_1)\cr
&& -\s_2^+(w_2){1\over 2 \pi i}{1\over 2 \pi i} \oint_{w_1}\oint_{w_1,|w-w_1|>|w'-w_1|}  G^-_{\dot{B}}(w) G^-_{\dot{A}}(w')\s_2^+(w_1)\diff w \diff w'\cr\cr
&& + \s_2^+(w_2){1\over 2 \pi i}\oint_{w_1} \diff w'G^-_{\dot{A}}(w')\s_2^+(w_1) {1\over 2 \pi i}\int_{\s =0,\t<\t_1}^{\s=2\pi}\diff w G^-_{\dot{B}}(w)
\eea
Here the extra sign change in the second term comes from swapping the order of $G^-_{\dot{A}}$ and $G^-_{\dot{B}}$. The contours  above $\s_2^+(w_2)$ and below $\s_{2}^+(w_1)$ wrap around the cylinder and have no power of $e^w$ multiplying $G^-_{\dot B}(w)$. Thus we can rewrite our expression as 
\bea
\hat{O}_{\dot{B}}\hat{O}_{\dot{A}} &=&\big( G^{(1),-}_{\dot{B},0} + G^{(2),-}_{\dot{B},0}\big)\s_2^+(w_2){1\over 2 \pi i}\oint_{w_1} \diff w'G^-_{\dot{A}}(w')\s_2^+(w_1)\cr
&& -\s_2^+(w_2){1\over 2 \pi i}{1\over 2 \pi i} \oint_{w_1}\oint_{w_1,|w-w_1|>|w'-w_1|}  G^-_{\dot{B}}(w) G^-_{\dot{A}}(w')\s_2^+(w_1)\diff w \diff w'\cr\cr
&& + \s_2^+(w_2){1\over 2 \pi i}\oint_{w_1} \diff w'G^-_{\dot{A}}(w')\s_2^+(w_1)\big( G^{(1),-}_{\dot{B},0} + G^{(2),-}_{\dot{B},0}\big)
\cr
\cr
&\equiv&\big( G^{(1),-}_{\dot{B},0} + G^{(2),-}_{\dot{B},0}\big)\mathcal{I}_1 - \mathcal{I}_2 + \mathcal{I}_{1}\big( G^{(1),-}_{\dot{B},0} + G^{(2),-}_{\dot{B},0}\big)
\label{full deformation one}
\eea
where in the last line we have made the following definitions
\bea
\mathcal{I}_1&\equiv& \s_2^+(w_2){1\over 2 \pi i}\oint_{w_1} \diff w G^-_{\dot{A}}(w)\s_2^+(w_1)\cr
\mathcal{I}_2&\equiv&\s_2^+(w_2) {1\over 2 \pi i}{1\over 2 \pi i} \oint_{w_1}\oint_{w_1,|w-w_1|>|w'-w_1|}  G^-_{\dot{B}}(w) G^-_{\dot{A}}(w')\s_2^+(w_1)\diff w \diff w'
\eea

We now have to evaluate the terms $\mathcal{I}_1$ and $\mathcal{I}_2$. This will be one of the main steps of our computation. We will evaluate each of these terms by mapping from the cylinder $w$ to the a covering space described by a coordinate  $t$. We now turn to this map.

\subsection{Mapping from the Cylinder to the $t$ Plane}

Let us now define the map from the cylinder to the $z$ plane:
\bea
z=e^{w}
\eea
The $z$ plane locations corresponding to initial and final copies of the CFT and the twist insertions are given by:
\bea
z&=&e^{-\infty+i\s}\to|z| = 0,~~~~\text{Copy 1 and 2 Final}\cr
z&=&e^{\infty+i\s}\to|z| = \infty,~~~~\text{Copy 1 and 2 Final}\cr
z_1&=&e^{w_1},~~~~ \text{First twist insertion}\cr
z_2&=&e^{w_2},~~~~ \text{Second twist insertion}
\label{w coor}
\eea
Mapping the measure combined with the supercharge as well as the twist operator to the $z$ plane gives
\bea
&&\diff w G^{-}_{\dot{A}}(w)\xrightarrow[]{w \to z} \diff z \bigg({\diff z\over \diff w} \bigg)^{1/2} G^{-}_{\dot{A}}(z) =\diff z z^{1/2} G^{-}_{\dot{A}}(z)\cr
&&\s^+_2(w_i)\xrightarrow[]{w \to z}\bigg({\diff z\over \diff w} \bigg)^{1/2}_{w=w_i}\s_2^+(z_i) =  z_i^{1/2}\s_2^{+}(z_i),\qquad i=1,2
\label{w to z}
\eea
Now let us map to the covering $t$ plane with the map:
\bea
z = {(t+a)(t+b)\over t}
\label{map}
\eea
Since our twist operators carry spin fields we will have to compute the images of the spin fields in the $t$ plane. These are bifurcation points. To do this we take the derivative of our map in (\ref{map}) and set it equal to zero:
\bea
{\diff z\over \diff t}={t^2-ab\over t^2}={(t-\sqrt{ab})(t+\sqrt{ab})\over t^2}=0
\eea
The solution to the above give the images of our spin fields to be:
\bea
t_1=-\sqrt{ab},\quad t_2 = \sqrt{ab} 
\label{twist image points}
\eea
Here we write the $t$ plane images corresponding to initial and final states on the cylinder where we use (\ref{w coor}):
\bea
z=0 \to  t&=&-a,~~~~ \text{Copy 1 Initial}\cr
z=0 \to t&=&-b ~~~~ \text{Copy 2 Initial}\cr
 z\to \infty \to t&=&\infty ~~~~ \text{Copy 1 Final}\cr
 z\to \infty\to t&=& 0  ~~~~ \text{Copy 2 Final}
\eea
where we have split the locations of the initial and final copies.
Now inserting the image points of our twist (\ref{twist image points}) into our map (\ref{map}), we define the twist insertion points in the $z$ plane in terms of the images of our initial copies $a$ and $b$:
\bea
z_1&\equiv&{(t_1+a)(t_1+b)\over t_1}=(\sqrt{a}-\sqrt{b})^2\cr
z_2&\equiv&{(t_2+a)(t_2+b)\over t_2} = (\sqrt{a} +\sqrt{b})^2
\label{coordinates}
\eea
We note that our the measure and supercharge transform as:
\bea
\diff z ~z^{1/2} G^{-}_{\dot{A}}(z)&\xrightarrow[]{z \to t}& \diff t (t+a)^{1/2}(t+b)^{1/2}t^{-1/2} \bigg({\diff z \over \diff t}\bigg)^{-1/2} G^{-}_{\dot{A}}(t)\cr
&=&  \diff t (t+a)^{1/2}(t+b)^{1/2}t^{-1/2} (t-t_1)^{-1/2}(t-t_2)^{-1/2}t G^{-}_{\dot{A}}(t)\cr
&=&  \diff t (t+a)^{1/2}(t+b)^{1/2} (t-t_1)^{-1/2}(t-t_2)^{-1/2}t^{1/2} G^{-}_{\dot{A}}(t)
\label{G coor map}
\eea
Let us define the separation of the twists as $\Delta w \equiv w_2 - w_1$. We can therefore define $w_1 = - {\Delta w \over 2}$ and $w_2 = {\Delta w \over 2}$. Using (\ref{w coor}), (\ref{coordinates}) and our definitions of $w_1$ and $w_2$, we can rewrite our $a$ and $b$ coordinates in terms of the twist separation $\Delta w$:
\bea
\sqrt{a} &=& {1\over 2}(\sqrt{z_2}+\sqrt{z_1})={1\over 2}\big(e^{{\Delta w \over 4
}} + e^{-{\Delta w \over 4}}\big)~~\to~~ a = \cosh^2\bigg({\Delta w \over 4 }\bigg)\cr
\sqrt{b} &=& {1\over 2}(\sqrt{z_2}-\sqrt{z_1})={1\over 2}\big(e^{{\Delta w \over 4}} - e^{-{\Delta w \over 4}}\big)~~\to~~ b = \sinh^2\bigg({\Delta w \over 4 }\bigg)
\eea 
We see that $a,b$ are strictly positive so we are confident to take the positive branch to be $\sqrt{ab}$ and the negative branch to be $-\sqrt{ab}$.
\subsection{Spectral Flows}
Since we started on the cylinder in the Ramond sector, the $t$ plane contains spin fields coming from the initial state (\ref{initial}), the final state given by $|\chi(w_1,w_2)\rangle$, and two twist insertions. In the final state $|\chi(w_1,w_2)\rangle$ we have an exponential of bilinear boson and fermion operators built on the vacuum $\rpmket$ where these bilinear operators are accompanied by bosonic and fermionic bogoluibov coefficients. To obtain a nonzero amplitude it is necessary to cap any given final state with the vacuum $\rpmbra$ which is given by:
\bea 
\rpmbra = \nsnsbra S^{(1)-}(\t=\infty)S^{(2)+}(\t = \infty)
\eea
We therefore see that this capping state also brings in spin fields.
We then remove all of the above mentioned spin fields by spectral flowing them away. Below, we record the spin fields sitting at finite points in the $t$ plane:
\bea
|0_R^-\rangle^{(1)} &\to& S^{-}(t=-a)\cr
|0_R^-\rangle^{(2)} &\to& S^{-}(t=-b)\cr
z_1^{1/2}\s_2^+(z_1)&\to&z_1^{1/2} S^{+}(t=t_1)\cr
z_2^{1/2}\s_2^+(z_2)&\to& z_2^{1/2} S^{+}(t=t_2)\cr
{}^{(2)}\langle 0_{R,-}| &\to& S^{+}(t=0)
\eea
We also have a spin field at $t=\infty$ coming from ${}^{(1)}\rpbra$. As we spectral flow away the spin fields at finite points the spin field at infinity will also be removed.
We will spectral flow in a way that does not introduce any new operators such as $J$'s. Under spectral flow, our $G^-$ changes as follows
\bea
G^{-}_{\dot{A}}(t) \to (t-t_i)^{ {\a \over 2}} G^{-}_{\dot{A}}(t)
\eea
where $t_i$ is the location of the spectral flow. Performing the following spectral flows
\bea
\a=+1 ~~\text{around}~~ t_i = -a,-b\cr
\a=-1 ~~\text{around}~~ t_i = t_1,t_2,0\cr
\eea
transforms our supercharge as follows:
\bea
G^-_{\dot{A}}(t)\to (t-a)^{1/2}(t-b)^{1/2}(t-t_1)^{-1/2}(t-t_2)^{-1/2}t^{-1/2}G^-_{\dot{A}}(t)
\label{sf}
\eea
Combining our spectral flow in (\ref{sf}) with the coordinate map from the $z$ to $t$ plane given in (\ref{G coor map}), our modification of the integrand and measure is given by:
\bea
 &&\diff t (t+a)^{1/2}(t+b)^{1/2} (t-t_1)^{-1/2}(t-t_2)^{-1/2}t^{1/2} G^{-}_{\dot{A}}(t)\cr
 &&\qquad\qquad \xrightarrow[]{t \to sf ~ t}  \diff t (t+a)(t+b) (t-t_1)^{-1}(t-t_2)^{-1}G^{-}_{\dot{A}}(t)\nn
\label{z to t sf transformation}
\eea
So the full transformation from $w$ to spectral flowed $t$ plane.
\bea
\diff w G^{-}_{\dot{A}}(w)&\xrightarrow[]{w \to z}& \diff z ~z^{1/2} G^{-}_{\dot{A}}(z)\cr
 &\xrightarrow[]{z \to t}&\diff t (t+a)^{1/2}(t+b)^{1/2} (t-t_1)^{-1/2}(t-t_2)^{-1/2}t^{1/2} G^{-}_{\dot{A}}(t)\cr
 &\xrightarrow[]{t \to sf ~ t}&   \diff t (t+a)(t+b) (t-t_1)^{-1}(t-t_2)^{-1}G^{-}_{\dot{A}}(t)\nn
\label{full G transformation}
\eea
We will also obtain an over factor of $C$ when performing our spectral flows coming from the presence of other spin fields. This constant depends upon the order in which the spectral flows were performed. We won't worry about the value of $C$ because we will invert the spectral flows in exactly the opposite order which will remove it.

Our next goal is to compute the amplitudes $\mathcal{I}_1$ and $\mathcal{I}_2$.

\section{Computing $\mathcal{I}_1$}\label{computing I1}

Now let us compute integral expression $\mathcal{I}_1$.  We start with the expression $\mathcal{I}_1$.  We apply the transformation (\ref{full G transformation}) and note  the Jacobian factor $(z_1z_2)^{1/2}$ coming from (\ref{w to z}).   We obtain
\bea
\mathcal{I}_1&\to& C(z_1z_2)^{1/2}{1\over 2 \pi i}\oint_{t_1} \diff t (t+a)(t+b) (t-t_1)^{-1}(t-t_2)^{-1}G^{-}_{\dot{A}}(t)
\label{I two}
\eea
We have removed all spin fields but we still have a singularity at $t=t_2$. To remove the singular behavior from $t_2$ we expand $(t-t_2)^{-1}$ around $t_1$:
\bea
(t-t_2)^{-1}=(t-t_1 + t_1 - t_2)^{-1}&=&\sum_{k=0}^\infty{}^{-1}C_{k}(t_1-t_2)^{-k-1}(t-t_1)^{k}\cr
&=&(t_1 - t_2)^{-1} + \sum_{k=1}^\infty{}^{-1}C_{k}(t_1-t_2)^{-k-1}(t-t_1)^{k}\nn
\label{expansion}
\eea
Inserting this into (\ref{I two}) we see that the only nonvanishing term be will the $k=0$ term because all others will annihilate the vacuum at $t_1$. Let us show this in more detail. We first define supercharge modes natural to the $t$ plane:
\bea
\tilde{G}^{\a,t\to t_1}_{\dot{A} , r}={1\over 2 \pi i}\oint \diff t (t-t_1)^{r+1/2} G^{\a}_{\dot{A},r} 
\label{t plane G mode}
\eea
Then we have
\bea
\tilde{G}^{-,t\to t_1}_{\dot{A} , r}|0_{NS}\rangle_{t_1} = 0,\quad r>-3/2
\eea
Therefore as we stated previously, the only $\tilde{G}^-_{\dot{A}}$ mode that survives when acting locally at $t_1$ is the one corresponding to $k=0$ in the expansion given in (\ref{expansion}) which is $\tilde{G}^{-,t\to t_1}_{\dot{A},-{3 \over 2}}$.
Therefore (\ref{I two}) becomes
\bea
\mathcal{I}_1\to C(z_1z_2)^{1/2}(t_1-t_2)^{-1}{1\over 2 \pi i}\oint_{t_2} \diff t (t+a)(t+b) (t-t_1)^{-1}G^{-}_{\dot{A}}(t)
\eea
We now only have contours around $t=t_1$.
We can now expand $\mathcal{I}_1$ to points corresponding to initial and final states in the $t$ plane.
\bea
\mathcal{I}_1 &\to& - C(z_1z_2)^{1/2}{1\over (t_2-t_1)}\cr
&&~\times~{1\over 2 \pi i} \bigg(\oint_{t=\infty} -\oint_{t=0} - \oint_{t=-a} -\oint_{=-b}  \bigg)\diff t (t+a)(t+b) (t-t_1)^{-1} G^{-}_{\dot{A}}(t)
\cr
\cr
\cr
&&=   C(z_1z_2)^{1/2}{1\over (t_2-t_1)}
\cr
&&\quad\times~{1\over 2 \pi i} \bigg(-\oint_{t=\infty} +  \oint_{t=0} + \oint_{t=-a} +  \oint_{=-b}  \bigg)\diff t (t+a)(t+b) (t-t_1)^{-1} G^{-}_{\dot{A}}(t)\nn
\label{I three}
\eea
where the minus signs come from the fact that contours at finite points reverse their order. Let us now undo the spectral flows. Performing spectral flows in the reverse direction, the integrand and measure change as follows 
\bea
  G^-_{\dot{A}}(t)\diff t &\xrightarrow[]{ t ~ \to ~ rsf ~t}& (t-a)^{-1/2}(t-b)^{-1/2}(t-t_1)^{1/2}(t-t_2)^{1/2}t^{1/2}G^-_{\dot{A}}(t)\diff t
  \cr
  \cr
 \one (t_2)\one(t_1)&\to& S^{+}(t_2)S^{+}(t_1)
 \label{t to rsf t}
\eea
where we have restored all spin fields but have only explicitly written those at the twist insertion points. 
Now let us map back to the $z$ plane. The combination of the measure with $G^-_{\dot{A}}(t)$ along with the spin fields transform as:
\bea
 G^-_{\dot{A}}(t)\diff t&\xrightarrow[]{rsf ~ t\to z}&\bigg({\diff z \over \diff t}\bigg)^{1/2}G^-_{\dot{A}}(z)\diff z =(t-t_1)^{1/2}(t-t_2)^{1/2}t^{-1}G^-_{\dot{A}}(z)\diff z\cr
\cr
 S^{+}(t_2)S^{+}(t_1) &\xrightarrow[]{rsf ~ t\to z}& \s_{2}^+(z_2)\s_{2}^+(z_1)
\label{rsf t to z}
\eea
Combining (\ref{t to rsf t}) and (\ref{rsf t to z}) with (\ref{I three}) gives $\mathcal{I}_1$ to be:
\bea
\mathcal{I}_1 \!\!\!\!&\to&\!\!\!\!   -(z_1z_2)^{1/2}{1\over (t_2-t_1)} \bigg[\bigg({1\over 2 \pi i}\oint_{z=\infty}\diff z ~z^{1/2}(t-t_2) G^{(1)-}_{\dot{A}}(t)\cr
&& + {1\over 2 \pi i}\oint_{z=\infty}\diff z~z^{1/2} (t-t_2) G^{(2)-}_{\dot{A}}(z)\bigg)\s_2^+(z_2)\s_2^+(z_1)\cr
&&\qquad\qquad -\s_2^+(z_2)\s_2^+(z_1)\bigg( {1\over 2 \pi i}\oint_{z=0}\diff z  ~ z^{1/2}(t-t_2) G^{(1)-}_{\dot{A}}(z)\cr
&& +  {1\over 2 \pi i}\oint_{z=0}\diff z ~ z^{1/2} (t-t_2) G^{(2)-}_{\dot{A}}(z)\bigg) \bigg]\nn
\eea
where we have used our map $z=(t+a)(t+b)t^{-1}$. We note the sign reversal for the integral $\oint_{t=0}$ because for $t=0$ the map goes like $z\sim {1\over t}$. This reverses the contour direction in the $z$ plane.
Now mapping this result back to the cylinder with the transformation
\bea
\diff z G^-_{\dot{A}}(z)&\xrightarrow[]{z \to w} & \diff w z^{-1/2} G^-_{\dot{A}}(w)
\cr
\cr
\s_{2}^{+}(z_2)\s_{2}^{+}(z_1)&\xrightarrow[]{z \to w} &(z_1z_2)^{-1/2}\s_{2}^{+}(w_2)\s_{2}^{+}(w_1)
\eea
gives $\mathcal{I}_1$ to be:
\bea
\mathcal{I}_1 \!\!\!\!&\to&\!\!\!\!   -{1\over (t_2-t_1)} \bigg[{1\over 2 \pi i}\int_{\s = 0,\t>\t_2}^{\s = 2\pi}\diff w (t-t_2)\bigg( G^{(1)-}_{\dot{A}}(w) + G^{(2)-}_{\dot{A}}(w) \bigg) \s_2^+(w_2)\s_2^+(w_1)\cr
&&\qquad\qquad -\s_2^+(w_2)\s_2^+(w_1){1\over 2 \pi i}\int_{\s = 0,\t<\t_1}^{\s = 2\pi}\diff w (t-t_2)\bigg( G^{(1)-}_{\dot{A}}(w) + G^{(2)-}_{\dot{A}}(w) \bigg) \bigg]\nn
\label{I four prime}
\eea
Let us take $t_1=-t_2$. Inserting this into (\ref{I four prime}) gives $\mathcal{I}_1$ to be:
\bea
\mathcal{I}_1 \!\!\!\!&\to&\!\!\!\!   -{1\over 2t_2} \bigg[{1\over 2 \pi i}\int_{\s = 0,\t>\t_2}^{\s = 2\pi}\diff w (t-t_2)\bigg( G^{(1)-}_{\dot{A}}(w) + G^{(2)-}_{\dot{A}}(w) \bigg) \s_2^+(w_2)\s_2^+(w_1)\cr
&&\qquad\qquad -\s_2^+(w_2)\s_2^+(w_1){1\over 2 \pi i}\int_{\s = 0,\t<\t_1}^{\s = 2\pi}\diff w (t-t_2)\bigg( G^{(1)-}_{\dot{A}}(w) + G^{(2)-}_{\dot{A}}(w) \bigg) \bigg]\nn
\label{I four}
\eea
Looking at (\ref{I four}), we see that we have written $\mathcal{I}_1$ in terms of $G^-_{\dot{A}}$ contours at initial and final states on the cylinder. Later we will write our contours in terms of cylinder modes where we will have to expand the $t$ coordinate in terms $z=e^w$. 

As a side note, let us summarize the total transformation going from the $t$ plane all the way back to the cylinder. We note that it is exactly opposite from the transformation (\ref{full G transformation}):
\bea
\diff t G^{-}_{\dot{A}}(t) &\xrightarrow[]{t\, \to rsf ~ t}&    (t+a)^{-1/2}(t+b)^{-1/2} (t-t_1)^{1/2}(t-t_2)^{1/2}t^{1/2} \diff t G^{-}_{\dot{A}}(t)\cr
 &\xrightarrow[]{rsf\, t\to z}&\diff z z^{-1/2} t^{-1}(t-t_1)(t-t_2) G^{-}_{\dot{A}}(z)\cr
&\xrightarrow[]{w \to z}& \diff w ~ (t+a)^{-1}(t+b)^{-1}(t-t_1)(t - t_2)G^{-}_{\dot{A}}(w)
\label{t to w}
\eea
and for the spin fields coming from twist insertions
\bea
\one (t_2)\one (t_1)&\xrightarrow[]{ t\to rsf~t}& S^{+}(t_2)S^{+}(t_1)\cr
S^{+}(t_2)S^{+}(t_1) &\xrightarrow[]{rsf ~ t\to z}& \s_{2}^+(z_2)\s_{2}^+(z_1)
\cr
\s_{2}^{+}(z_2)\s_{2}^{+}(z_1)&\xrightarrow[]{z \to w} &(z_1z_2)^{-1/2}\s_{2}^{+}(w_2)\s_{2}^{+}(w_1)
\label{spin field t to w}
\eea
In the next section we compute the amplitude $\mathcal{I}_2$ following the same procedure as we did here for $\mathcal{I}_1$.

\section{Computing $\mathcal{I}_2$}\label{computing I2}

Now let us compute the integral expression $\mathcal{I}_2$. We start with the cylinder expression:
\bea
\mathcal{I}_2&\equiv&\s_2^+(w_2) {1\over 2 \pi i}{1\over 2 \pi i} \oint_{w_1}\oint_{w_1,|w-w_1|>|w'-w_1|}  G^-_{\dot{B}}(w) G^-_{\dot{A}}(w')\s_2^+(w_1)\diff w \diff w'
\eea
Let us map from the cylinder to the $t$ plane and perform the necessary spectral flows. In doing this we obtain a modification in our integral expression that is similar to $\mathcal{I}_{1}$ except now we have two integrals instead of one. Therefore inserting the result in (\ref{full G transformation}) for both the $t$ integral and the $t'$ integral gives
\bea
\mathcal{I}_{2} &\to& CC'(z_1z_2)^{1/2}{1\over 2 \pi i}{1\over 2 \pi i} \oint_{t_1}\oint_{t_1,|t-t_1|>|t'-t_1|}(t+a)(t+b) (t-t_1)^{-1}(t-t_2)^{-1} \cr
&&(t'+a)(t'+b) (t'-t_1)^{-1}(t'-t_2)^{-1} G^-_{\dot{B}}(t) G^-_{\dot{A}}(t')\diff t \diff t'
\label{GG one prime}
\eea
where the constants $C$ and $C'$ come from spectral flows in both the $t$ and $t'$ coordinates respectively. Both constants will again be removed later when spectral flowing in exactly the reverse order.
We again remove singular terms at $t_2$ by expanding them around $t_1$. Inserting the expansion given in (\ref{expansion}) into (\ref{GG one prime}) for both $(t-t_2)^{-1}$ and $(t'-t_2)^{-1}$ gives $\mathcal{I}_2$ to be:
\bea
\mathcal{I}_{2} &\equiv& CC'(z_1z_2)^{1/2}\sum_{k=0}^{\infty}(-1)^k(t_1-t_2)^{-k-2} {1\over 2 \pi i}{1\over 2 \pi i} \oint_{t_1}\oint_{t_1,|t-t_1|>|t'-t_1|}(t+a)(t+b)\cr
&& (t-t_1)^{k-1} (t'+a)(t'+b) (t'-t_1)^{-1} G^-_{\dot{B}}(t) G^-_{\dot{A}}(t')\diff t \diff t'
\label{GG one}
\eea
Here we make several comments about the nontrivial terms from our $k$ expansions.
Since the $\dot{A}$ contour is on the inside the only term which survives the $(t'-t_2)^{-1}$ expansion is the $k=0$ term which, as shown when computing $\mathcal{I}_1$, is $\tilde{G}^{-,t\to t_1}_{\dot{A},-{3\over 2}}$. For the $\dot{B}$ contour however we will have the $k=0$ term plus additional terms corresponding to $k=1,2$ in the expansion of $(t-t_2)^{-1}$. These additional terms anticommute nontrivially with the $\dot{A}$ contour to produce $\tilde{J}^-$ terms. This can be seen when looking at the anticommutation relation:
\bea
\nn
\ac{\tilde{G}^{\alpha,t\to t_1}_{\dot{A},m}}{\tilde{G}^{\beta,t\to t_1}_{\dot{B},n}} &=& \hspace*{-4pt}\epsilon_{\dot{A}\dot{B}}\bigg[
   (m^2 - \frac{1}{4})\epsilon^{\alpha\beta}\delta_{m+n,0}
  + (m-n){(\sigma^{aT})^\alpha}_\gamma\epsilon^{\gamma\beta}\tilde{J}^{a,t\to t_1}_{m+n}
  + \epsilon^{\alpha\beta} \tilde{L}^{t\to t_1}_{m+n}\bigg]\quad\nn
  \label{anticommutator}
\eea 
where $\tilde{G}^{\a,t\to t_1}_{\dot{A},m}$ is defined in (\ref{t plane G mode}).
The $t$ plane $J^{-}$ terms are defined locally at $t_1$ as:
\bea
\tilde{J}^{-,t\to t_1}_{n} = {1\over 2 \pi i}\oint_{t_1}\diff t (t-t_1)^n J^-(t)
\label{t plane J}
\eea
However, instead of actually anticommuting the $\dot{B}$ mode for $k=1,2$ to produce $\tilde{J}^-$ terms, we leave our expression in terms of $\tilde{G}^-$'s. The reason is as follows: when we did anticommute the $\dot{B}$ modes through the $\dot{A}$ mode we obtained terms that were proportional to $\tilde{J}^{-,t\to t_1}_{-1}$ and $\tilde{J}^{-,t\to t_1}_{-2}$ with $J^{-,t\to t_1}_n$ defined in (\ref{t plane J}). We found that for the $\tilde{J}^{-,t\to t_1}_{-1}$ term, the cylinder expansion was trivial because the cylinder contour carried no factor of $(t-t_2)$ which appears in $\mathcal{I}_1$ expression. However, for the $\tilde{J}^{-,t\to t_1}_{-2}$ term, the cylinder contour carried a factor $(t-t_2)^{-1}$ which becomes extremely difficult to expand in terms of the coordinate $z=e^w$ which is necessary in order to write our two deformation operator completely in terms of cylinder modes. To avoid this problem altogether we simply leave our integral expression, $\mathcal{I}_2$, in terms of $G^-$ contours. 

We note here that we have two cases to consider for $\mathcal{I}_2$ which we examine below. The first case is when $\dot{A}=\dot{B}$ and the second case is for general $\dot{A}$ and $\dot{B}$. We note that we compute the special case of $\dot{A}=\dot{B}$ in the $t$ plane as opposed to back on the cylinder.
\subsection{\underline{$\dot{A}=\dot{B}$}}\label{A = B one}
We first begin with the case of  $\dot{A}=\dot{B}$. With this constraint the expression $\mathcal{I}_2$ can be written as:
\bea
\mathcal{I}_{2} &\equiv& CC'(z_1z_2)^{1/2}\sum_{k=0}^{2}(-1)^k(t_1-t_2)^{-k-2} {1\over 2 \pi i}{1\over 2 \pi i} \oint_{t_1}\oint_{t_1,|t-t_1|>|t'-t_1|}(t+a)(t+b)\cr
&& (t-t_1)^{k-1} (t'+a)(t'+b) (t'-t_1)^{-1} G^-_{\dot{A}}(t) G^-_{\dot{A}}(t')\diff t \diff t'
\label{GG two}
\eea
Let us insert the $t$ plane mode expansions given in (\ref{t plane G mode}) into (\ref{GG two}).
Therefore (\ref{GG two}) becomes
\bea
\mathcal{I}_{2} &\to&  CC'(z_1z_2)^{1/2}(t_1-t_2)^{-2}(t_1+a)^2(t_1+b)^2 \tilde{G}^{-,t\to t_1}_{\dot{A},-3/2} \tilde{G}^{-,t\to t_1}_{\dot{A},-3/2}
\cr
\cr
&&\quad - CC'(z_1z_2)^{1/2}(t_1-t_2)^{-3} (t_1 + a)^2(t_1+b)^2\tilde{G}^{-,t\to t_1}_{\dot{A},-1/2} \tilde{G}^{-,t\to t_1}_{\dot{A},-3/2}
\cr
\cr
&&\quad + CC'(z_1z_2)^{1/2}(t_1-t_2)^{-4} (t_1+a)^2(t_1+b)^2\tilde{G}^{-,t\to t_1}_{\dot{A},1/2} \tilde{G}^{-,t\to t_1}_{\dot{A},-3/2}
\eea
Since we have a local $NS$ vacuum at $t_1$ we compute the action of each term on the vacuum. Looking at the first term we find:
\bea
\tilde{G}^{t\to t_1,-}_{\dot{A},-3/2}\tilde{G}^{t\to t_1,-}_{\dot{A},-3/2}|0_{NS}\rangle_{t_1}=0
\eea
since two fermions aren't allowed to be in the same state. Looking at the second and third terms we find:
\bea
&& \tilde{G}^{-,t\to t_1}_{\dot{A},-1/2}\tilde{G}^{-,t\to t_1}_{\dot{A},-3/2}|0_{NS}\rangle_{t_1}=-\tilde{G}^{-,t\to t_1}_{\dot{A},-3/2}\tilde{G}^{-,t\to t_1}_{\dot{A},-1/2}|0_{NS}\rangle_{t_1}=0\cr
&&  \tilde{G}^{-,t\to t_1}_{\dot{A},1/2}\tilde{G}^{-,t\to t_1}_{\dot{A},-3/2}|0_{NS}\rangle_{t_1}=-\tilde{G}^{-,t\to t_1}_{\dot{A},-3/2}\tilde{G}^{-,t\to t_1}_{\dot{A},1/2}|0_{NS}\rangle_{t_1}=0
\eea
where we have used anticommutation relation (\ref{anticommutator}). We see that these terms also vanish and therefore conclude that for $\dot{A} = \dot{B}$ the integral expression $\mathcal{I}_2$ vanishes. This gives a drastic simplification to the full expression of the two deformation operators given in (\ref{full deformation one}).

In the next section we compute the integral expression for $\mathcal{I}_2$ for general $\dot{A}$ and $\dot{B}$.
\subsection{\underline{General $\dot{A}$ and $\dot{B}$}}
We now compute integral expression for $\mathcal{I}_2$ for general $\dot{A}$ and $\dot{B}$. Again writing our expression we have:
\bea
 \mathcal{I}_{2}&\to&CC'(z_1z_2)^{1/2}\sum_{k=0}^{2}{}^{-1}C_{k}(t_1-t_2)^{-k-2} {1\over 2 \pi i}{1\over 2 \pi i} \oint_{t_1}\oint_{t_1,|t-t_1|>|t'-t_1|}(t+a)(t+b)\cr
&& (t-t_1)^{k-1} (t'+a)(t'+b) (t'-t_1)^{-1} G^-_{\dot{B}}(t) G^-_{\dot{A}}(t')\diff t \diff t'
\label{spectral flow expanded GG}
\eea
Taking:
\bea
{}^{-1}C_{k}=(-1)^k
\eea 
gives:
\bea
 \mathcal{I}_{2} &\equiv&  CC'(z_1z_2)^{1/2}\sum_{k=0}^{2}(-1)^k(t_1-t_2)^{-k-2} {1\over 2 \pi i}{1\over 2 \pi i} \oint_{t_1}\oint_{t_1,|t-t_1|>|t'-t_1|}(t+a)(t+b)\cr
&& (t-t_1)^{k-1} (t'+a)(t'+b) (t'-t_1)^{-1} G^-_{\dot{B}}(t) G^-_{\dot{A}}(t')\diff t \diff t'
\eea

Let us now expand both the $t$ and $t'$ contours away from $t_1$ to points corresponding to initial and final points on the cylinder. Since each contour can land on four different points stretching them will produce a total of sixteen terms. However, we must take care to keep up with minus signs as contours will be reversing direction at finite points. Let us tabulate the minus sign changes. We note that stretching the $\dot{B}$ contour to finite points will reverse direction as well as place it on the inside of the $\dot{A}$ contour that also lands on that point however we always leave the $\dot{B}$ supercharge outside of the $\dot{A}$ supercharge. When the $\dot{B}$ and $\dot{A}$ contour both land at $t=\infty$ we get no sign change because the contours are in the right direction and the and $\dot{A}$ contour is still inside of the $\dot{B}$ contour. Proceeding forward, we have
\bea
&&\dot{B}~~\text{at}~~\infty,\quad\dot{A}~~\text{at} ~~\infty   \quad \to \quad (+1)(+1)=+1\cr
&&\dot{B}~~\text{at}~~\infty,\quad\dot{A}~~\text{at} ~~0 \quad \to \quad (+1)(-1)=-1\cr
&&\dot{B}~~\text{at}~~\infty,\quad\dot{A}~~\text{at} ~~a   \quad \to \quad (+1)(-1)=-1\cr
&&\dot{B}~~\text{at}~~\infty,\quad\dot{A}~~\text{at} ~~b   \quad \to \quad (+1)(-1)=-1\cr
\cr
\cr
&&\dot{B}~~\text{at}~~ 0,\quad\dot{A}~~\text{at} ~~\infty   \quad \to \quad (-1)(+1)=-1\cr
&&\dot{B}~~\text{at}~~ 0,\quad\dot{A}~~\text{at} ~~0   \quad \to \quad (-1)(-1)=+1\cr
&&\dot{B}~~\text{at}~~ 0,\quad\dot{A}~~\text{at} ~~-a   \quad \to \quad (-1)(-1)=+1\cr
&&\dot{B}~~\text{at}~~ 0,\quad\dot{A}~~\text{at} ~~-b   \quad \to \quad  (-1)(-1)=+1\cr
\cr
&&\dot{B}~~\text{at}~~ -a,\quad\dot{A}~~\text{at} ~~\infty   \quad \to \quad  (-1)(+1)=-1\cr
&&\dot{B}~~\text{at}~~ -a,\quad\dot{A}~~\text{at} ~~0   \quad \to \quad  (-1)(-1)=+1\cr
&&\dot{B}~~\text{at}~~ -a,\quad\dot{A}~~\text{at} ~~-a   \quad \to \quad  (-1)(-1)=+1\cr
&&\dot{B}~~\text{at}~~ -a,\quad\dot{A}~~\text{at} ~~-b   \quad \to \quad  (-1)(-1)=+1\cr
\cr
\cr
&&\dot{B}~~\text{at}~~ -b,\quad\dot{A}~~\text{at} ~~\infty   \quad \to \quad  (-1)(+1)=-1\cr
&&\dot{B}~~\text{at}~~ -b,\quad\dot{A}~~\text{at} ~~0   \quad \to \quad  (-1)(-1)=+1\cr
&&\dot{B}~~\text{at}~~ -b,\quad\dot{A}~~\text{at} ~~-a   \quad \to \quad  (-1)(-1)=+1\cr
&&\dot{B}~~\text{at}~~ -b,\quad\dot{A}~~\text{at} ~~-b   \quad \to \quad  (-1)(-1)=+1
\eea
Implementing these sign changes as we stretch our contours in (\ref{spectral flow expanded GG}) gives the following:
\bea
 \mathcal{I}_{2}&\to&CC'(z_1z_2)^{1/2}\sum_{k=0}^{2}(-1)^k(t_1-t_2)^{-k-2}\cr
&& {1\over 2 \pi i}{1\over 2 \pi i} \bigg(\oint_{t=\infty}\oint_{t'=\infty, |t|>|t'|} - \oint_{t=\infty}\oint_{t'=0} - \oint_{t=\infty}\oint_{t'=-a} - \oint_{t=\infty}\oint_{t'=-b}\cr
&& - \oint_{t=0}\oint_{t'=\infty} + \oint_{t=0}\oint_{t'=0,|t'|>|t|}  + \oint_{t=0}\oint_{t'=-a} + \oint_{t=0}\oint_{t'=-b} \cr
&& - \oint_{t=-a}\oint_{t'=\infty} + \oint_{t=-a}\oint_{t'=0}  + \oint_{t'=-a} \oint_{t=-a,|t'+a|>|t+a|}   +  \oint_{t=-a}\oint_{t'=-b}\cr
&&  - \oint_{t=-b}\oint_{t'=\infty} + \oint_{t=-b}\oint_{t'=0} + \oint_{t=-b}\oint_{t'=-a}  + \oint_{t'=-b}\oint_{t=-b,|t'+b|>|t+b|}  \bigg )\cr
\cr
&&\qquad (t+a)(t+b) (t-t_1)^{k-1}(t'+a)(t'+b) (t'-t_1)^{-1} G^-_{\dot{B}}(t) G^-_{\dot{A}}(t')\diff t \diff t'
\nn
\label{expanded GG} 
\eea
Let us rearrange (\ref{expanded GG}) into three terms grouped according to the contour locations specified below:
\begin{enumerate}
\item  $t=0,\infty; ~~\quad t'=0,\infty$\\
\item  $t=0,\infty;~~\quad t'=-a,-b~~~~$ and $~~\quad t=-a,-b;\quad t'=0,\infty$\\
\item  $t=-a,-b; ~~\quad t'=-a,-b$ 
\end{enumerate}
Contours circling points  in set 1 are placed on the cylinder  after the two twists. For set 2, each term has one contour before the twists and one after the twists. For set 3, the contours are both placed before the twists. 
 
Therefore (\ref{expanded GG}) becomes:
\bea
 \mathcal{I}_{2}&\to&CC'(z_1z_2)^{1/2}\sum_{k=0}^{2}(-1)^k(t_1-t_2)^{-k-2}\cr
&& {1\over 2 \pi i}{1\over 2 \pi i} \bigg(\oint_{t=\infty}\oint_{t'=\infty, |t|>|t'|} - \oint_{t=\infty}\oint_{t'=0}  - \oint_{t=0}\oint_{t'=\infty} + \oint_{t=0}\oint_{t'=0,|t'|>|t|}\bigg)
\cr
\cr
&&\qquad\qquad\times~(t+a)(t+b) (t-t_1)^{k-1}(t'+a)(t'+b) (t'-t_1)^{-1} G^-_{\dot{B}}(t) G^-_{\dot{A}}(t')\diff t \diff t'
\cr
\cr
&&~ +  ~CC'(z_1z_2)^{1/2}\sum_{k=0}^{2}(-1)^k(t_1-t_2)^{-k-2}\cr
&& \quad\times~{1\over 2 \pi i}{1\over 2 \pi i}\bigg( - \oint_{t=\infty}\oint_{t'=-a} - \oint_{t=\infty}\oint_{t'=-b} + \oint_{t=0}\oint_{t'=-a} + \oint_{t=0}\oint_{t'=-b}\cr
&&\qquad\qquad  - \oint_{t=-a}\oint_{t'=\infty} + \oint_{t=-a}\oint_{t'=0} - \oint_{t=-b}\oint_{t'=\infty} + \oint_{t=-b}\oint_{t'=0}\bigg)
\cr
\cr
&&\qquad\qquad\quad\times~(t+a)(t+b) (t-t_1)^{k-1}(t'+a)(t'+b) (t'-t_1)^{-1} G^-_{\dot{B}}(t) G^-_{\dot{A}}(t')\diff t \diff t'
\cr
\cr
&& ~ + ~ CC'(z_1z_2)^{1/2}\sum_{k=0}^{2}(-1)^k(t_1-t_2)^{-k-2}\cr
&&\quad\times~ {1\over 2 \pi i}{1\over 2 \pi i}\bigg(\oint_{t'=-a} \oint_{t=-a,|t'+a|>|t+a|}   + \oint_{t=-a}\oint_{t'=-b} \cr
&&\qquad\qquad\qquad ~ +  \oint_{t=-b}\oint_{t'=-a} + \oint_{t'=-b}\oint_{t=-b,|t'+b|>|t+b|} \bigg ) (t+a)(t+b) (t-t_1)^{k-1}\cr
&&\qquad\qquad\qquad\qquad\times~(t'+a)(t'+b) (t'-t_1)^{-1} G^-_{\dot{B}}(t) G^-_{\dot{A}}(t')\diff t \diff t' 
\label{expanded GG rearranged prime}
\eea

Let us now reverse spectral flow in exactly the opposite order in order to remove the constants $C$ and $C'$ and then map back to the the cylinder. Even though we don't explicitly write the intermediate step of mapping from the $z$ plane to the $t$ plane we note that there is sign change that occurs for any integral around $t$ or $t'=0$. This is again because in mapping from the $t$ plane to the $z$ plane the contours around $t=0$ reverse the direction of the contour bringing in a minus sign just as they did when mapping from the $z$ plane to the $t$ plane. In the case where both integrals are around $t=t'=0$, we will gain two minus signs for that term giving no overall sign change. We note that the $\dot{B}$ contour maps from inside of the $\dot{A}$ contour in the $t$ plane, to outside the $\dot{A}$ contour in the $z$ plane. Then the same ordering is kept when going to the cylinder. Using the transformation in (\ref{t to w}) and (\ref{spin field t to w}) for both $G$ contours gives the following cylinder expression for $\mathcal{I}_2$:
\bea
 \mathcal{I}_{2}&\to&\sum_{k=0}^{2}(-1)^k(t_1-t_2)^{-k-2}\cr
&&\times~\bigg[{1\over 2 \pi i}\oint_{\s=0,\t>\t_2}^{\s=2\pi} \diff w~ (t-t_1)^k(t-t_2) \bigg( G^{(1)-}_{\dot{B}}(w) +  G^{(2)-}_{\dot{B}}(w)\bigg)\cr
&&\qquad\times{1\over 2 \pi i}\oint_{\s'=0,\t'>\t_2,\t>\t'}^{\s'=2\pi} \diff w'~ (t'-t_2) \bigg( G^{(1)-}_{\dot{A}}(w') +  G^{(2)-}_{\dot{A}}(w')\bigg)\bigg]\s^+_2(w_2)\s^+_2(w_1)
\cr
\cr
\cr
&&+\sum_{k=0}^{2}(-1)^k(t_1-t_2)^{-k-2}\cr
&&\quad\times~\bigg[ - {1\over 2 \pi i}\oint_{\s=0,\t>\t_2}^{\s=2\pi}\diff w~ (t-t_1)^k(t-t_2)\bigg( G^{(1)-}_{\dot{B}}(w) + G^{(2)-}_{\dot{B}}(w)\bigg)\s^+_2(w_2)\s^+_2(w_1)\cr
&&\quad\quad\quad\quad \times{1\over 2 \pi i}\oint_{\s'=0,\t'<\t_1}^{\s'=2\pi} \diff w'~ (t'-t_2)\bigg( G^{(1)-}_{\dot{A}}(w') + G^{(2)-}_{\dot{A}}(w')\bigg)
\cr\cr
&&\quad\quad\quad{}+ {1\over 2 \pi i}\oint_{\s'=0,\t'>\t_2}^{\s'=2\pi}\diff w'~ (t'-t_2)\bigg( G^{(1)-}_{\dot{A}}(w') + G^{(2)-}_{\dot{A}}(w')\bigg)\s^+_2(w_2)\s^+_2(w_1)\cr
&&\quad\quad\quad\quad\times ~ {1\over 2 \pi i}\oint_{\s=0,\t<\t_1}^{\s=2\pi} \diff w~ (t-t_1)^k(t-t_2)\bigg( G^{(1)-}_{\dot{B}}(w) + G^{(2)-}_{\dot{B}}(w)\bigg)\bigg]\cr
\cr
\cr
&&- \sum_{k=0}^{2}(-1)^k(t_1-t_2)^{-k-2} \s^+_2(w_2)\s^+_2(w_1)\cr
&&\quad\times ~\bigg[{1\over 2 \pi i}\oint_{\s'=0,\t'<\t_1}^{\s'=2\pi} \diff w'~ (t'-t_2) \bigg( G^{(1)-}_{\dot{A}}(w') +  G^{(2)-}_{\dot{A}}(w')\bigg)\cr
&&\quad\quad\times~{1\over 2 \pi i}\oint_{\s=0,\t<\t_1,\t'>\t}^{\s=2\pi} \diff w~ (t-t_1)^k(t-t_2) \bigg( G^{(1)-}_{\dot{B}}(w) +  G^{(2)-}_{\dot{B}}(w)\bigg)\bigg]\nn
\label{expanded GG rsf z to w}
\eea

We have now computed the integral expression $\mathcal{I}_2$ in terms of contours before and after the twists on the cylinder. Our next goal is to write the full expression of the deformation operator on the cylinder before expanding our coordinate $t$ in terms of the cylinder coordinate $w$.

We again remind the reader that the general case reduces to the case where $\dot{A}=\dot{B}$ that we also computed. However, it is much easier to evaluate this while still in the $t$ plane as we have done as opposed to waiting until we have reached the cylinder.

\section{Full expression before expanding $t$ in terms of cylinder coordinate $w$}\label{cylinder result}

Here we compute the full deformation operator expressions on the cylinder prior to expanding our coordinate $t$ in terms of $z=e^w$. We compute the two cases below. 

\subsection{\underline{$\dot{A}=\dot{B}$}}

Looking at the case when $\dot{A}=\dot{B}$, in (\ref{A = B one}) we found that $\mathcal{I}_2=0$ which gives a great simplification. Taking $\mathcal{I}_2=0$ in (\ref{full deformation one}) gives the following expression for the two deformation operator:
\bea
\hat{O}_{\dot{A}}\hat{O}_{\dot{A}}&\equiv& \big( G^{(1),-}_{\dot{B},0} + G^{(2),-}_{\dot{B},0}\big)\mathcal{I}_1 + \mathcal{I}_{1}\big( G^{(1),-}_{\dot{B},0} + G^{(2),-}_{\dot{B},0}\big)
\label{A = B cylinder expression}
\eea
Now inserting our expression for $\mathcal{I}_1$ which is given in ( \ref{I four}) into (\ref{A = B cylinder expression}) gives us the following:
\bea
\hat{O}_{\dot{A}}\hat{O}_{\dot{A}}\!\!{}&\equiv&{}\!\!- {1\over 2t_2}\big( G^{(1),-}_{\dot{A},0} + G^{(2),-}_{\dot{A},0}\big)\cr
&&\quad\bigg[\bigg( {1\over 2 \pi i}\int_{\s = 0,\t>\t_2}^{\s = 2\pi}\diff w (t-t_2)G^{(1)-}_{\dot{A}}(w)\cr
&&\qquad\qquad + ~ {1\over 2 \pi i}\int_{\s = 0,\t>\t_2}^{\s = 2\pi}\diff w (t-t_2)G^{(2)-}_{\dot{A}}(w) \bigg) \s_2^+(w_2)\s_2^+(w_1)
\cr
\cr
&&\quad~-\s_2^+(w_2)\s_2^+(w_1) ~ \bigg( {1\over 2 \pi i}\int_{\s = 0,\t<\t_1}^{\s = 2\pi}\diff w (t-t_2)G^{(1)-}_{\dot{A}}(w)\cr
&&\qquad\qquad\qquad\qquad\qquad\qquad + {1\over 2 \pi i}\int_{\s = 0,\t<\t_1}^{\s = 2\pi}\diff w (t-t_2)G^{(2)-}_{\dot{A}}(w) \bigg) \bigg]
\cr
\cr
\cr
&&  -{1\over 2t_2}\bigg[\bigg({1\over 2 \pi i}\int_{\s = 0,\t>\t_2}^{\s = 2\pi}\diff w (t-t_2) G^{(1)-}_{\dot{A}}(w)\cr
&&\qquad\qquad+ ~{1\over 2 \pi i}\int_{\s = 0,\t>\t_2}^{\s = 2\pi}\diff w (t-t_2)G^{(2)-}_{\dot{A}}(w) \bigg) ~ \s_2^+(w_2)\s_2^+(w_1)\cr
&&\qquad~~ -~\s_2^+(w_2)\s_2^+(w_1)\bigg( {1\over 2 \pi i}\int_{\s = 0,\t<\t_1}^{\s = 2\pi}\diff w (t-t_2)G^{(1)-}_{\dot{A}}(w)\cr
&&\qquad\qquad\qquad\qquad\qquad\qquad {}+ {1\over 2 \pi i}\int_{\s = 0,\t<\t_1}^{\s = 2\pi}\diff w (t-t_2)G^{(2)-}_{\dot{A}}(w) \bigg) \bigg]\cr
&&\quad\times\big( G^{(1),-}_{\dot{B},0} + G^{(2),-}_{\dot{B},0}\big)
\label{A = B two}
\eea
This is our cylinder expression for the two deformation operators for the case where $\dot{A}=\dot{B}$ before computing the $G^-$ contour integrals in terms of cylinder modes. Next we record the cylinder expression for our two deformation operator for general $\dot{A}$ and $\dot{B}$.
\subsection{\underline{General $\dot{A}$ and $\dot{B}$}}
Now looking at the case for general $\dot{A}$ and $\dot{B}$ we insert (\ref{expanded GG rsf z to w}) and (\ref{I four}) into (\ref{full deformation one}) giving the expression:
\bea
\hat{O}_{\dot{B}}\hat{O}_{\dot{A}}\!\! &=&\!\!  -{1\over  2t_2}\big( G^{(1),-}_{\dot{B},0} + G^{(2),-}_{\dot{B},0}\big)\cr
&&\times~\bigg[ \bigg({1\over 2 \pi i}\int_{\s = 0,\t>\t_2}^{\s = 2\pi}\diff w (t-t_2) G^{(1)-}_{\dot{A}}(w)\cr
&&\quad\quad{} + {1\over 2 \pi i}\int_{\s = 0,\t>\t_2}^{\s = 2\pi}\diff w (t-t_2) G^{(2)-}_{\dot{A}}(w) \bigg)\s_2^+(w_2)\s_2^+(w_1)
\cr
\cr
\cr
&&~ -\s_2^+(w_2)\s_2^+(w_1)\bigg({1\over 2 \pi i}\int_{\s = 0,\t<\t_1}^{\s = 2\pi}\diff w (t-t_2) G^{(1)-}_{\dot{A}}(w)\cr
&&\qquad\qquad\qquad\qquad\qquad{} + {1\over 2 \pi i}\int_{\s = 0,\t<\t_1}^{\s = 2\pi}\diff w (t-t_2)G^{(2)-}_{\dot{A}}(w) \bigg) \bigg]
\cr
\cr
\cr
&&\!\!\!\!\!\! -\sum_{k=0}^{2}(-1)^k(t_1-t_2)^{-k-2}\cr
&&\times~\bigg[\bigg({1\over 2 \pi i}\int_{\s=0,\t>\t_2}^{\s=2\pi} \diff w~ (t-t_1)^k(t-t_2) G^{(1)-}_{\dot{B}}(w)
\cr
&&\quad\quad +  {1\over 2 \pi i}\int_{\s=0,\t>\t_2}^{\s=2\pi} \diff w~(t-t_1)^k (t-t_2)G^{(2)-}_{\dot{B}}(w)\bigg)
\cr
&&\quad\times\bigg({1\over 2 \pi i}\int_{\s'=0,\t'>\t_2,\t>\t'}^{\s'=2\pi} \diff w'~ (t'-t_2) G^{(1)-}_{\dot{A}}(w')\cr
&&\quad\quad{}+  {1\over 2 \pi i}\int_{\s'=0,\t'>\t_2,\t>\t'}^{\s'=2\pi} \diff w'~ (t'-t_2)G^{(2)-}_{\dot{A}}(w')\bigg)\bigg]\s^+_2(w_2)\s^+_2(w_1)
\cr
\cr
\cr
&&\!\!\!\!\!\!+ \sum_{k=0}^{2}(-1)^k(t_1-t_2)^{-k-2}\cr
&&\times~\bigg[  \bigg({1\over 2 \pi i}\oint_{\s=0,\t>\t_2}^{\s=2\pi}\diff w~ (t-t_1)^k(t-t_2) G^{(1)-}_{\dot{B}}(w)
\cr
&&\quad\quad {}+  {1\over 2 \pi i}\oint_{\s=0,\t>\t_2}^{\s=2\pi}\diff w~ (t-t_1)^k(t-t_2)G^{(2)-}_{\dot{B}}(w)\bigg)\s^+_2(w_2)\s^+_2(w_1)
\cr
\cr
&&\quad\times\bigg( {1\over 2 \pi i}\int_{\s'=0,\t'<\t_1}^{\s'=2\pi} \diff w'~ (t'-t_2) G^{(1)-}_{\dot{A}}(w')\cr
&&\quad\quad{}+  {1\over 2 \pi i}\int_{\s'=0,\t'<\t_1}^{\s'=2\pi} \diff w'~ (t'-t_2)G^{(2)-}_{\dot{A}}(w')\bigg)
\cr
\cr
\cr
&&\quad - \bigg({1\over 2 \pi i}\int_{\s'=0,\t'>\t_2}^{\s'=2\pi}\diff w'~ (t'-t_2) G^{(1)-}_{\dot{A}}(w')\cr
&&\quad\quad{} + {1\over 2 \pi i}\int_{\s'=0,\t'>\t_2}^{\s'=2\pi}\diff w'~ (t'-t_2) G^{(2)-}_{\dot{A}}(w')\bigg)\s^+_2(w_2)\s^+_2(w_1)
\cr
&&\quad\quad \times\bigg({1\over 2 \pi i}\int_{\s=2,\pi\t<\t_1}^{\s=2\pi} \diff w~ (t-t_1)^k(t-t_2) G^{(1)-}_{\dot{B}}(w)
\cr
&&\quad\quad\quad+ {1\over 2 \pi i}\int_{\s=2,\pi\t<\t_1}^{\s=2\pi} \diff w~ (t-t_1)^k(t-t_2) G^{(2)-}_{\dot{B}}(w)\bigg)\bigg]
\cr
\cr
\cr
&&\!\!\!\!\!\!+\s^+_2(w_2)\s^+_2(w_1) \sum_{k=0}^{2}(-1)^k(t_1-t_2)^{-k-2}\cr
&&\times~\bigg[\bigg({1\over 2 \pi i}\int_{\s'=0,\t'<\t_1}^{\s'=2\pi} \diff w'~ (t'-t_2)  G^{(1)-}_{\dot{A}}(w')\cr
&&\quad\quad{} + {1\over 2 \pi i}\int_{\s'=0,\t'<\t_1}^{\s'=2\pi} \diff w'~ (t'-t_2) G^{(2)-}_{\dot{A}}(w')\bigg)\cr
&&\quad\times\bigg({1\over 2 \pi i}\int_{\s=0,\t<\t_1,\t'>\t}^{\s=2\pi} \diff w~ (t-t_1)^k(t-t_2) G^{(1)-}_{\dot{B}}(w)
\cr
&&\quad\quad {} +  {1\over 2 \pi i}\int_{\s=0,\t<\t_1,\t'>\t}^{\s=2\pi}\diff w(t-t_1)^k(t-t_2)G^{(2)-}_{\dot{B}}(w)\bigg)\bigg]
\cr
\cr
\cr
&&\!\!\!\!\!\!  -{1\over 2t_2}\bigg[\bigg({1\over 2 \pi i}\int_{\s = 0,\t>\t_2}^{\s = 2\pi}\diff w (t-t_2) G^{(1)-}_{\dot{A}}(w)\cr
&&\quad\quad{} + {1\over 2 \pi i}\int_{\s = 0,\t>\t_2}^{\s = 2\pi}\diff w (t-t_2)G^{(2)-}_{\dot{A}}(w) \bigg)\s_2^+(w_2)\s_2^+(w_1)
\cr
\cr
&&\quad~~ -~\s_2^+(w_2)\s_2^+(w_1)\bigg( {1\over 2 \pi i}\int_{\s = 0,\t<\t_1}^{\s = 2\pi}\diff w (t-t_2)G^{(1)-}_{\dot{A}}(w)\cr
&&\qquad\qquad\qquad\qquad\qquad\qquad{} + {1\over 2 \pi i}\int_{\s = 0,\t<\t_1}^{\s = 2\pi}\diff w (t-t_2)G^{(2)-}_{\dot{A}}(w) \bigg) \bigg]\cr
&&\quad\times\big ( G^{(1),-}_{\dot{B},0} + G^{(2),-}_{\dot{B},0} \big )
\label{full deformation two}
\eea

We have now written the expression for the two deformation operators on the cylinder for both the specific case $\dot{A} = \dot{B}$ and the more general case $\dot{A}\neq\dot{B}$. Remarkably, we were able to write the expression in terms of $G$ contours before and after the twists. Our next step is to expand the coordinate $t$ in terms of $z=e^w$ in order to write the full deformation operator in terms of cylinder modes.  

\section{Expanding $t$ in terms of the cylinder coordinate $w$}\label{map inversion}

Now that we have written our two deformation operator as contours on the cylinder our goal is to write these contours in terms of cylinder modes.

\subsection{Map Inversion}

To do this we must first invert our map so that we can expand the integrands in terms of the cylinder coordinate $e^{w}$. We start with the following:
\bea
&&z={(t+a)(t+b)\over t}\cr
&&\to zt= t^2+(a+b)t +ab\cr
&& \to t^2+t(a+b-z) +ab = 0
\eea
Inserting this result into \textit{Mathematica} and using $t_2=\sqrt{ab}$ gives the solutions
\bea
t_{\pm}&=&{1\over 2}\bigg(z-(a+b)\pm \sqrt{-(2t_2)^2 + (a+b-z)^2}\bigg)\cr
&=&{1\over 2}\bigg(z-(a+b)\pm \sqrt{(z - (a+b+2\sqrt{ab}))( z - (a+b-2\sqrt{ab}))}\bigg)\nn
\label{t solution}
\eea
Let us now look at the limiting behavior in order to determine which solutions will correspond to which initial and final states.

\subsubsection*{ Final Copies $\t\to\infty\Rightarrow z\to \infty$}

Taking the limit as $z\to\infty$ we have:
\bea
t_{\pm}={1\over 2}(z \pm \sqrt{z^2})
\eea
Taking the positive branch of $ \sqrt{z^2}$:
\bea
t_{\pm} = {1\over 2}(z \pm z)
\eea
We there see that the $t_+$ solution corresponds to Copy $1$ final because 
\bea
t\sim z
\eea
and similarly $t_-$ corresponds to Copy $2$ final because
\bea
t\sim 0
\eea
We also perform this same analysis for the initial copies:
\subsubsection*{ Initial Copies $\t \to -\infty \Rightarrow z\to 0$}
Taking the limit $z\to 0$
\bea
t_{\pm}&=&{1\over 2}\bigg(-(a+b)\pm \sqrt{(b-a)^2}\bigg)
\eea
Staying consistent with the final copy notation we want the $+$ solution to correspond to Copy 1 initial. We therefore take the positive branch of $\sqrt{(b-a)^2}$ we get:
\bea
t_{\pm}&=&{1\over 2}\bigg(-(a+b)\pm (b-a)\bigg)
\eea
The $t_+$ therefore corresponds to Copy $1$ because:
\bea
t\sim-a
\eea
and the $t_-$ solution corresponds to Copy $2$ because
\bea
t\sim -b
\eea
Now that we have picked the appropriate solutions for the appropriate initial and final states we proceed forward with evaluating $t_{\pm}$. Substituting
\bea
(a+b+2\sqrt{ab})&=&z_2=e^{{\Delta w\over 2}}\nn
(a+b-2\sqrt{ab})&=&z_1=e^{-{\Delta w\over 2}}\nn
a+b &=& -2t_2 + e^{{\Delta w\over 2}}
\eea
into (\ref{t solution}) gives
\bea
t_{\pm}&=&{1\over 2}\bigg(z-(a+b)\pm \sqrt{(z - e^{{\Delta w \over 2}})(z - e^{-{\Delta w \over 2}})}\bigg)
\label{t pm}
\eea
%Let us reduce 
%\bea
 %-2t_2 + e^{{\Delta w\over 2}}&=&-2\sqrt{ab} + e^{{\Delta w\over 2}}\cr
 %&=& -2\cosh\bigg({\Delta w\over 4 }\bigg)\sinh\bigg({\Delta w\over 4 }\bigg) + %e^{{\Delta w\over 2}}\cr
% &=& -\sinh\bigg({\Delta w\over 2 }\bigg) + e^{{\Delta w\over 2}}\cr
% &=& -\sinh\bigg({\Delta w\over 2 }\bigg) + \cosh\bigg({\Delta w\over 2 }\bigg) +  %%\sinh\bigg({\Delta w\over 2 }\bigg)\cr
% &=&\cosh\left({\Delta w\over 2 }\right)
 %\label{trig reduction}
%\eea
%Implementing these changes into our solution we have:
%\bea
%t_{\pm} & = &{1\over 2} \bigg( z - \bigg(e^{{\Delta w\over 2}} - 2t_2  \bigg)\bigg)\pm  {1\over 2}\big(z - e^{{\Delta w \over 2}}\big)^{1/2}\big(z - e^{-{\Delta w \over 2}}\big)^{1/2}
%
%\eea
Since we will have to expand $z$ around different points corresponding to initial and final states, let us do each one in turn in the following subsections. 

\subsection{Copy 1 and Copy 2 Final}

Here we expand our $t$ coordinate around $\t = \infty\to z =\infty$ in order to write our $G$ contours in terms of Copy 1 and Copy 2 final modes on the cylinder. Looking at the square root terms in (\ref{t pm}), we have a general expansion of the form 
\bea
\big(z+x\big)^{1/2} &=& z^{1/2}\big(1+xz^{-1}\big)^{1/2}\cr
&=&  z^{1/2}\big(1+xz^{-1}\big)^{1/2}\cr
&=& \sum_{p_1\geq 0}{}^{1/2}C_{p_1} z^{-p_1 + 1/2}x^{p_1}
\eea
Applying the above expansion to (\ref{t pm}) we obtain
\bea
t_{\pm} = {1\over 2} ( z - (a+b)) \pm   {1\over 2}\sum_{p_1\geq 0}\sum_{p_2\geq 0}{}^{1/2}C_{p_1}{}^{1/2}C_{p_2} z^{-p_1-p_2 + 1}(-1)^{p_1+p_2}e^{{\Delta w \over 2}(p_1-p_2)}
\label{solution}
\eea
which are the solutions, $t_{\pm}$, at final points on the cylinder. Let us make the following variable redefinitions:
\bea
k&=&{p_1 - p_2\over 2}\cr
n &=& p_1 + p_2 \cr
\Rightarrow p_1&=& {n\over 2}+k\cr
\Rightarrow p_2&=&{n\over 2} - k
\label{redefinition}
\eea
Evaluating the limits give:
\bea
&& p_1\geq 0 ~~\Rightarrow~~ k+ {n\over 2}\geq 0 ~~\Rightarrow k~~ \geq -  {n\over 2}
\cr
&& p_2\geq 0 ~~\Rightarrow~~  {n\over 2}-k\geq 0 ~~\Rightarrow {n\over 2} \geq k
\label{limits}
\eea
our solutions, (\ref{solution}), become:
\bea
t_{\pm} = {1\over 2} ( z - (a+b)) \pm   {1\over 2}\sum_{n = 0}^{\infty} (-1)^{n} z^{-n + 1}\sum_{k = - {n\over 2}}^{{n \over 2}}{}^{1/2}C_{{n\over 2} + k}{}^{1/2}C_{{n\over 2} - k}e^{k\Delta w}
\label{solution two}
\eea
Looking at our $k$ sum we see for each $k = j$ term  with $j\in \mathbb{Z}_+$ there will be a corresponding $k =-j $ which allows us to make the following replacement:
\bea
e^{k\Delta w}\to \cosh\big( k\Delta w \big)
\label{cosine argument}
\eea
Applying this replacement to (\ref{solution two}) gives:
\bea
t_{\pm} &=& {1\over 2} ( z - (a+b) ) \pm   {1\over 2}\sum_{n = 0}^{\infty} (-1)^{n} z^{-n + 1}\sum_{k = - {n\over 2}}^{{n \over 2}}{}^{1/2}C_{{n\over 2} + k}{}^{1/2}C_{{n\over 2} - k}\cosh\big(k \Delta w\big)\cr
&=& {1\over 2} \big( z - (a+b) \pm  \sum_{n = 0}^{\infty}C_n  z^{1-n}\big)
\label{t plus z}
\eea
where we've made the following definition:
\bea
C_n&\equiv&  (-1)^{n}\sum_{k = - {n\over 2}}^{{n \over 2}}{}^{1/2}C_{{n\over 2} + k}{}^{1/2}C_{{n\over 2} - k}\cosh\big( k \Delta w\big)
\label{C}
\eea
Making the replacement $z=e^w$ in (\ref{t plus z}) gives
\bea
t_{\pm}&=&  {1\over 2} \big( e^w - (a+b) \pm  \sum_{n = 0}^{\infty}C_n  e^{w(1-n)}\big)
\label{t plus}
\eea

Here we expanded our $t$ coordinate in terms of the coordinate $z=e^w$ for Copy 1 and Copy 2 final modes. Next we will expand the $t$ coordinate around Copy 1 and Copy 2 initial states.
 
\subsection{Copy 1 and Copy 2 Initial}

Here we expand our $t$ coordinate around $\t = -\infty\to z =0$. Looking back at $t_{\pm}$ in (\ref{t pm}) we must expand the square root portion around around small $z$. We therefore have:
\bea
\big(z+x\big)^{1/2} &=& x^{1/2}\big(1+x^{-1}z \big)^{1/2}\cr
 &=&\sum_{p_1\geq 0}{}^{1/2}C_{p_1}x^{-p_1 +1/2}z^{p_1}
\eea
Applying this to (\ref{t pm}) gives:
\bea
t_{\pm} &=& {1\over 2} ( z -(a+b)) \mp {1\over 2}\sum_{p_1\geq 0}\sum_{p_2\geq 0}{}^{1/2}C_{p_1}{}^{1/2}C_{p_2} z^{p_1+p_2 }(-1)^{p_1+p_2}e^{-{\Delta w \over 2}(p_1-p_2)}\nn
\label{t pm initial}
\eea
Again making the substitutions given in (\ref{redefinition}) and (\ref{limits}), (\ref{t pm initial}) becomes:
\bea
t_{\pm} = {1\over 2} ( z - (a+b )) \mp  {1\over 2}\sum_{n = 0}^{\infty} (-1)^{n} z^{ n }\sum_{k = - {n\over 2}}^{{n \over 2}}{}^{1/2}C_{{n\over 2} + k}{}^{1/2}C_{{n\over 2} - k}e^{-k\Delta w}
\label{t pm initial two}
\eea
Using the same argument given in (\ref{cosine argument}) we write (\ref{t pm initial two}) as:
\bea
t_{\pm} &=& {1\over 2} \big( z -( a+b ) \mp \sum_{n = 0}^{\infty} C_n z^{ n }\big)\nn
\label{t pm one initial z}
\eea
Again inserting $z=e^w$ in (\ref{t pm one initial z}) gives:
\bea
t_{\pm} &=& {1\over 2} \big( e^w -( a+b ) \mp \sum_{n = 0}^{\infty} C_n e^{ nw }\big)\nn
\label{t pm one initial}
\eea
We see that the only difference between the $t_{\pm}$ expansion for initial and final copies comes in the \textbf{third} term. There is a modification of the mode number as well as a minus sign difference. We summarize this difference below:
\bea
&&\text{Final Copies}: \pm \sum_{n = 0}^{\infty} C_n e^{ (1-n)w }\cr
&&\text{Initial Copies} :\mp \sum_{n = 0}^{\infty} C_n e^{ nw }
\label{third term modification}
\eea

We have now written our $t$ coordinate in terms of expansions around the location of Copy 1 and Copy 2 initial states. 

In the final section we compute the integral expressions on the cylinder by inserting the expansions that we have computed for Copy 1 and Copy 2 final as well as Copy 1 and Copy 2 initial. We then compute the expression for the final result of the two deformation operators on the cylinder.

\section{Computing Supercharge Contours on the Cylinder}\label{cylinder modes} 

Our goal in this section is to compute the expression of the full two deformation operator in terms of cylinder modes using the expansions computed in Section \ref{map inversion}. We first define the supercharge modes on the cylinder. We then compute the necessary contour integrals in terms of cylinder modes at locations corresponding to Copy 1 and Copy 2 final states as well as Copy 1 and Copy 2 initial states that appear in our two deformation operators. 

\subsection{Supercharge modes on the cylinder}

Here we define the supercharge modes on the cylinder. We will have modes before the two twists and modes after the two twists. Our modes after the twist are defined as follows:
\subsubsection*{\underline{$\t>\t_2$}}
\bea
G^{(1),-}_{\dot{C},n}\equiv {1\over 2 \pi i}\int_{\s = 0,\t > \t_2}^{2\pi }  \diff w G^{(1),-}_{\dot{C}}(w) e^{nw}\cr
G^{(2),-}_{\dot{C},n}\equiv {1\over 2 \pi i}\int_{\s = 0,\t > \t_2}^{2\pi }  \diff w G^{(2),-}_{\dot{C}}(w) e^{nw}\cr
\label{cylinder G}
\eea
with $n$ being at integer.
Our modes before the twist are defined as follows:
\subsubsection*{\underline{$\t<\t_1$}}
\bea
G^{(1),-}_{\dot{C},n}\equiv{1\over 2 \pi i}\int_{\s = 0,\t < \t_1}^{2\pi }  \diff w G^{(1),-}_{\dot{C}}(w) e^{nw}\cr
G^{(2),-}_{\dot{C},n}\equiv{1\over 2 \pi i}\int_{\s = 0,\t < \t_1}^{2\pi }  \diff w G^{(2),-}_{\dot{C}}(w) e^{nw}\cr
\label{cylinder G initial}
\eea
Here we have defined our supercharge modes on the cylinder. Next we compute the contour integrals on the cylinder in terms of these cylinder modes.
\subsection{Computing Copy 1 and Copy 2 Final Contours}
Now we compute the necessary contours that appear in (\ref{A = B two}) and (\ref{full deformation two}) in terms of cylinder modes defined in (\ref{cylinder G}). There will be contours of two types in these expressions. We compute the simpler contours first.
\subsubsection{Contours containing $(t-t_2)$} 
The first contours we compute are of the form:
\bea
\text{Copy 1}&:&{1\over 2 \pi i}\int_{\s = 0,\t>\t_2}^{\s = 2\pi}\diff w (t_+-t_2) G^{(1)-}_{\dot{C}}(w)
\cr
\cr
\cr
\text{Copy 2}&:&{1\over 2 \pi i}\int_{\s = 0,\t>\t_2}^{\s = 2\pi}\diff w (t_--t_2) G^{(2)-}_{\dot{C}}(w)
\label{t minus t2 final prime}
\eea
where $t_{\pm}$ correspond to the two roots that were computed in Section \ref{map inversion} with $t_+$ corresponding to Copy 1 and $t_-$ corresponding to Copy 2. Using (\ref{t plus}) we compute the following contribution:
\bea
t_{\pm} - t_2 &=&  {1\over 2} \big( e^w - (a+b) \pm  \sum_{n = 0}^{\infty}C_n  e^{(1-n)w}\big)-t_2\cr
&=&  {1\over 2} \big( e^w - e^{\Delta w \over 2} \pm  \sum_{n = 0}^{\infty}C_n  e^{(1-n)w}\big)
\label{t minus tpm}
\eea
Inserting (\ref{t minus tpm}) into (\ref{t minus t2 final prime}) and using the mode definitions in (\ref{cylinder G}) gives:
\bea
\text{Copy 1}&:&{1\over 2 \pi i}\int_{\s = 0,\t>\t_2}^{\s = 2\pi}\diff w (t_+-t_2) G^{(1)-}_{\dot{C}}(w) \cr
&&\qquad = {1\over 2}\bigg ( G^{(1)-}_{\dot{C},1} - e^{\Delta w\over 2}G^{(1)-}_{\dot{C},0}  +  \sum_{n = 0}^{\infty}  C_n G^{(1)-}_{\dot{C},1-n} \bigg) \cr
\cr
\cr
\text{Copy 2}&:&{1\over 2 \pi i}\int_{\s = 0,\t>\t_2}^{\s = 2\pi}\diff w (t_--t_2) G^{(2)-}_{\dot{C}}(w)\cr
&&\qquad = {1\over 2}\bigg ( G^{(2)-}_{\dot{C},1} - e^{\Delta w\over 2}G^{(2)-}_{\dot{C},0}  -  \sum_{n = 0}^{\infty}  C_n G^{(2)-}_{\dot{C},1-n} \bigg) 
\label{t minus t2 final}
\eea
Let us define the symmetric antisymmtric basis:
\bea
G^{-}_{\dot{C},n} &=&  \big( G^{(1)-}_{\dot{C},n} +  G^{(2)-}_{\dot{C},n}  \big )\cr
\tilde{G}^{-}_{\dot{C},n} &=& \big( G^{(1)-}_{\dot{C},n} - G^{(2)-}_{\dot{C},n}\big)
\label{new basis}
\eea
We do this because computing amplitudes will turn out to be easier when we write all of our modes in terms of the symmetric and antisymmetric basis.

Now combining both results given in (\ref{t minus t2 final}) and using the basis (\ref{new basis}) gives the following expression:
\bea
&&{1\over 2 \pi i}\int_{\s = 0,\t>\t_2}^{\s = 2\pi}\diff w (t_+-t_2) G^{(1)-}_{\dot{C}}(w) + {1\over 2 \pi i}\int_{\s = 0,\t>\t_2}^{\s = 2\pi}\diff w (t_--t_2) G^{(2)-}_{\dot{C}}(w)\cr
&&\qquad =  {1\over 2}\bigg ( G^{-}_{\dot{C},1} - e^{\Delta w\over 2}G^{-}_{\dot{C},0}  + \sum_{n = 0}^{\infty}  C_n \tilde{G}^{-}_{\dot{C},1-n} \bigg)
\label{t minus t2 final 2}
\eea
For this type of term we see a finite number of symmetric annihilation modes \textit{after} the twist and an infinite number of antisymmetric creation modes \textit{after} the twists. Since creation modes annihilate when acting on a state to the left we need not worry about getting an infinite number of terms.
\subsubsection{Contours containing $\sum_{k=0}^{2}(t_1-t_2)^{-k-2}(-1)^k(t-t_2)(t-t_1)^k$} 
Now we compute the more complicated contours of the form:
\bea
\text{Copy 1}&:&{1\over 2 \pi i}\int_{\s = 0,\t>\t_2}^{\s = 2\pi}\diff w \sum_{k=0}^{2}(t_1-t_2)^{-k-2}(-1)^k(t_+-t_2)(t_+-t_1)^k G^{(1),-}_{\dot{C}}(w)
\cr
\cr
\text{Copy 2}&:&{1\over 2 \pi i}\int_{\s = 0,\t>\t_2}^{\s = 2\pi}\diff w \sum_{k=0}^{2}(t_1-t_2)^{-k-2}(-1)^k(t_--t_2)(t_--t_1)^k G^{(2),-}_{\dot{C}}(w)
\label{Copy 1 final}
\eea
Let us simplify the following term that we get coming in the integrand of (\ref{Copy 1 final}):
\bea
&&\sum_{k=0}^{2}(-1)^{k}(t_1-t_2)^{-k-2}(t_{\pm}-t_2)(t_{\pm}-t_1)^k\cr
&&\qquad = {1\over 4 t_2^2}\bigg(  (t_{\pm}-t_2) - {(t_{\pm}-t_2)(t_{\pm}+t_2) \over -2 t_2} +   {(t_{\pm}-t_2)(t_{\pm}+t_2)^2 \over 4t_2^2}  \bigg)\cr
&&\qquad = {1\over 16 t_2^4}\bigg( (t_{\pm}-t_2)(t_{\pm}^2 + 4 t_{\pm} t_2 + 7 t_2^2)  \bigg)\cr
&&\qquad = {1\over 16 t_2^4}\bigg(t_{\pm} (z^2 - z A + B)-zt_2^2 + z_1t_2^2-8t_2^3\bigg)
\label{expansion of integrand}
\eea
where considerable simplification was made in (\ref{expansion of integrand}). We've also made the following definitions:
\bea
A &\equiv&(2z_1 + t_2)\cr
B &\equiv& z_1 (a+b)+ t_2(z_2 - 2(a+b))
\label{A B}
\eea
We note that we used the relation
\bea
z={(t+a)(t+b)\over t}\to t^2=(z-(a+b))t -t_2^2
\eea
to reduce powers of $t$ in (\ref{expansion of integrand}).
First inserting $z=e^w$ into the RHS of (\ref{expansion of integrand}) and then using, (\ref{t plus}) we write (\ref{expansion of integrand}) as: 
\bea
&&\sum_{k=0}^{2}(-1)^{k}(t_1-t_2)^{-k-2}(t_{\pm}-t_2)(t_{\pm}-t_1)^k\cr
&&\qquad =  {1\over 32 t_2^4}\bigg( e^{3w}-e^{2w}\big(A + a+ b\big)+e^w\big(A(a+b)+ B - 2t_2^2 \big)\cr
&&\qquad\qquad \pm\bigg(\sum_{n = 0}^{\infty}C_n  e^{(3-n)w} -   A\sum_{n = 0}^{\infty}C_n  e^{(2-n)w} + B \sum_{n = 0}^{\infty}C_n  e^{(1-n)w}\bigg)\cr
&&\qquad\qquad +\big(2z_1t_2^2 - (a+b)B - 16 t_2^3 \big)\bigg) \nn
\label{expansion of integrand 2}
\eea
Here we have written our expression in terms of the cylinder coordinate $w$. Simplifying (\ref{expansion of integrand 2}) using the definitions of $A$ and $B$ in (\ref{A B}) give:
\bea
&&\sum_{k=0}^{2}(-1)^{k}(t_1-t_2)^{-k-2}(t_{\pm}-t_2)(t_{\pm}-t_1)^k\cr
&&\qquad =  {1\over 32  t_2^4}\bigg[ e^{3w} - A'e^{2w} + B'e^w + C'\cr
&&\qquad\qquad\qquad\qquad \pm \bigg(  \sum_{n = 0}^{\infty}C_n  e^{(3-n)w} -  A\sum_{n = 0}^{\infty}C_n  e^{(2-n)w} +B \sum_{n = 0}^{\infty}C_n  e^{(1-n)w}\bigg)\bigg]
\nn
\label{tpm one}
\eea
where we make the following definitions:
\bea
A'&\equiv&A+a+b\cr
B'&\equiv&A(a+b) + B - 2t_2^2\cr
C'&\equiv& 2z_1t_2^2 -  (a+b)B - 16 t_2^3
\label{A B C prime}
\eea
Let us combine the sums together. To do this let us expand each individual sum in (\ref{tpm one}) appropriately:
\bea
\sum_{n = 0}^{\infty}C_n  e^{(3-n)w} &=&  C_0e^{3w} +  C_1e^{2w} +\sum_{n = 2}^{\infty}C_n e^{(3-n)w}\cr
&=& e^{3w}  -(a+b)e^{2w} +\sum_{n = 2}^{\infty}C_n e^{(3-n)w}
\cr
\cr
A\sum_{n = 0}^{\infty}C_n e^{(2-n)} &=& A e^{2w}  +   A\sum_{n = 1}^{\infty}C_n e^{(2-n)w}
\label{sum post twist}
\eea
Now shifting each sum back to $n=0$ gives:

\bea
\sum_{n = 0}^{\infty}C_n  e^{(3-n)w} &=& e^{3w}  -(a+b)e^{2w} + \sum_{n = 0}^{\infty}C_{n+2} e^{(1-n)w}
\cr
\cr
A\sum_{n = 0}^{\infty}C_n e^{(2-n)w} &=& A e^{2w} + A\sum_{n = 0}^{\infty}C_{n+1} e^{(1-n)w}   
\label{sum two post twist}
\eea
Inserting (\ref{sum two post twist}) into term containing the three summations in (\ref{tpm one}) gives:
\bea
&&  \sum_{n = 0}^{\infty}C_n  e^{(3-n)w} -  A\sum_{n = 0}^{\infty}C_n  e^{(2-n)w} +B \sum_{n = 0}^{\infty}C_n  e^{(1-n)w} 
 \cr
 \cr
 &&\qquad\qquad =  e^{3w} - A'e^{2w} + \sum_{n=0}^{\infty} \big (C_{n+2}- A C_{n+1} + B C_n\big)e^{(1-n)w}
 \cr
 \cr
  &&\qquad\qquad=  e^{3w} - A'e^{2w} + \sum_{n=0}^{\infty}C'_n e^{(1-n)w}
  \label{sum}
\eea
where
\bea
C'_n\equiv (C_{n+2}- A C_{n+1} + B C_n\big)
\label{Cn prime}
\eea
Inserting (\ref{sum}) into  (\ref{tpm one}) gives:
\bea
&&\sum_{k=0}^{2}(-1)^{k}(t_1-t_2)^{-k-2}(t_{\pm}-t_2)(t_{\pm}-t_1)^k\cr
&&\quad\quad =  {1\over 32 t_2^4}\bigg( e^{3w} - A'e^{2w} + B'e^w + C' \pm \bigg( e^{3w} - A'e^{2w} + \sum_{n=0}^{\infty}C'_n e^{(1-n)w}\bigg)\bigg)  
\label{t pm two}
\eea
We have now reduced our term to one infinite sum.
Now inserting (\ref{t pm two}) into (\ref{Copy 1 final}) for both Copy 1 and Copy 2 contours, taking $t=t_+$ for Copy 1 and $t=t_-$ for Copy 2, and using the appropriate mode definitions given in (\ref{cylinder G}), we obtain the following result for the Copy 1 and Copy 2 final contours:
\bea
\text{Copy } 1&:&{1\over 2 \pi i}\int_{\s = 0,\t>\t_2}^{\s = 2\pi}\diff w \sum_{k=0}^{2}(-1)^k(t_1-t_2)^{-k-2}(t_+-t_2)(t_+-t_1)^k G^{(1)-}_{\dot{C}}(w)
\cr
\cr
&&\qquad\qquad = {1\over 32 t_2^4}\bigg[ G_{\dot{C},3}^{(1)-}-A'G_{\dot{C},2}^{(1)-}+B'G_{\dot{C},1}^{(1)-} + C'  G_{\dot{C},0}^{(1)-}\cr
&&\qquad\qquad\qquad\qquad\qquad\qquad{} + \bigg( G_{\dot{C},3}^{(1)-} - A'G_{\dot{C},2}^{(1)-} + \sum_{n=0}^{\infty}C'_n G_{\dot{C},1-n}^{(1)-}\bigg)\bigg] 
\cr
\cr
\cr
\cr
\text{Copy } 2&:&{1\over 2 \pi i}\int_{\s = 0,\t>\t_2}^{\s = 2\pi}\diff w \sum_{k=0}^{2}(-1)^k(t_1-t_2)^{-k-2}(t_--t_2)(t_--t_1)^k G^{(2)-}_{\dot{C}}(w)
\cr
\cr
&&\qquad\qquad = {1\over  32 t_2^4}\bigg[ G_{\dot{C},3}^{(2)-}-A'G_{\dot{C},2}^{(2)-}+B'G_{\dot{C},1}^{(2)-} + C'  G_{\dot{C},0}^{(2)-}\cr
&&\qquad\qquad\qquad\qquad\qquad\qquad{} - \bigg( G_{\dot{C},3}^{(2)-} - A'G_{\dot{C},2}^{(2)-} + \sum_{n=0}^{\infty}C'_n G_{\dot{C},1-n}^{(2)-}\bigg)\bigg]\nn  
\label{cylinder contours 1}
\eea
Using our mode definitions in (\ref{new basis}) we add together both Copy 1 and Copy 2 contours given in (\ref{cylinder contours 1}) to obtain:
\bea
&&{1\over 2 \pi i}\int_{\s = 0,\t>\t_2}^{\s = 2\pi}\diff w \sum_{k=0}^{2}(-1)^k(t_1-t_2)^{-k-2}(t_+-t_2)(t_+-t_1)^k G^{(1)-}_{\dot{C}}(w)
\cr
&& \quad + 
{1\over 2 \pi i}\int_{\s = 0,\t>\t_2}^{\s = 2\pi}\diff w \sum_{k=0}^{2}(-1)^k(t_1-t_2)^{-k-2}(t_--t_2)(t_--t_1)^k G^{(2)-}_{\dot{C}}(w)
\cr
\cr
&&\quad = {1\over  32 t_2^4}\bigg( G_{\dot{C},3}^{-} - A'G_{\dot{C},2}^{-} + B'G_{\dot{C},1}^{-} + C'  G_{\dot{C},0}^{-} +  \tilde{G}_{\dot{C},3}^{-} - A'\tilde{G}_{\dot{C},2}^{-} + \sum_{n=0}^{\infty}C'_n \tilde{G}_{\dot{C},1-n}^{-}\bigg)\nn
\label{combined final}
\eea
We note that there are a finite number of both symmetric and antisymmetric annihilation modes \textit{after} the twist and an infinite number of creation modes \textit{after} the twist. Again, since creation modes annihilate when acting on a state to the left we need not worry about getting an infinite number of terms.

We have now computed the two relevant types contours at locations on the cylinder corresponding to Copy 1 and Copy 2 \textit{final} states. Next, we compute these same contours at locations on the cylinder corresponding to Copy 1 and Copy 2 \textit{initial} states.

\subsection{Integral Expressions for Copy 1 and Copy 2 Initial}

Here we compute the relevant contours corresponding to Copy 1 and Copy 2 \textit{initial} states. We will again have two types of terms analogous to the case for Copy 1 and Copy 2 \textit{final} states.

\subsubsection{Contours containing $(t-t_2)$} 

Just as before we begin with the simpler of the two contours with terms of the following form:
\bea
\text{Copy 1}&:&{1\over 2 \pi i}\int_{\s = 0,\t<\t_1}^{\s = 2\pi}\diff w (t_+-t_2) G^{(1)-}_{\dot{C}}(w)
\cr
\cr
\cr
\text{Copy 2}&:&{1\over 2 \pi i}\int_{\s = 0,\t<\t_1}^{\s = 2\pi}\diff w (t_--t_2) G^{(2)-}_{\dot{C}}(w)
\label{t minus t2 initial prime}
\eea
where again we take $t_+$ to correspond to Copy 1 initial and $t_1$ to correspond to Copy 2 initial.
For these contours we compute the following expression using (\ref{t pm one initial}):
\bea
t_{\pm} - t_2 &=&  {1\over 2} \big( e^w - e^{\Delta w \over 2} \mp  \sum_{n = 0}^{\infty}C_n  e^{nw}\big)
\label{t minus tpm initial}
\eea
Inserting (\ref{t minus tpm initial}) into (\ref{t minus t2 initial prime}) and using the mode definitions given in (\ref{cylinder G initial}), we obtain the following result for Copy 1 and Copy 2 initial contours:
\bea
\text{Copy 1}&:&{1\over 2 \pi i}\int_{\s = 0,\t<\t_1}^{\s = 2\pi}\diff w (t_+-t_2) G^{(1)-}_{\dot{C}}(w) \cr
&&\qquad = {1\over 2}\bigg ( G^{(1)-}_{\dot{C},1} - e^{\Delta w\over 2}G^{(1)-}_{\dot{C},0}  -  \sum_{n = 0}^{\infty}  C_n G^{(1)-}_{\dot{C},n} \bigg) \cr
\cr
\cr
\cr
\text{Copy 2}&:&{1\over 2 \pi i}\int_{\s = 0,\t<\t_1}^{\s = 2\pi}\diff w (t_--t_2) G^{(2)-}_{\dot{C}}(w)\cr
&&\qquad = {1\over 2}\bigg ( G^{(2)-}_{\dot{C},1} - e^{\Delta w\over 2}G^{(2)-}_{\dot{C},0}  +  \sum_{n = 0}^{\infty}  C_n G^{(2)-}_{\dot{C},n} \bigg) 
\label{t minus t2 initial}
\eea
Combining both contours given in (\ref{t minus t2 initial}) and using the basis defined in (\ref{new basis}) gives:
\bea
&&{1\over 2 \pi i}\int_{\s = 0,\t<\t_1}^{\s = 2\pi}\diff w (t_+-t_2) G^{(1)-}_{\dot{C}}(w) + {1\over 2 \pi i}\int_{\s = 0,\t<\t_1}^{\s = 2\pi}\diff w (t_--t_2) G^{(2)-}_{\dot{C}}(w)\cr
&&\qquad =  {1\over 2}\bigg ( G^{-}_{\dot{C},1} - e^{\Delta w\over 2}G^{-}_{\dot{C},0}  -  \sum_{n = 0}^{\infty}  C_n \tilde{G}^{-}_{\dot{C},n} \bigg)
\label{t minus t2 initial 2}
\eea
We note here that we have a finite number of symmetric and antisymmetric annihilation modes \textit{before} the twist and infinite number of antisymmetric annihilation modes \textit{before} the twist. Since annihilation modes annihilate when acting on a state to the right we need not worry about getting an infinite number of terms.

Here we have now computed the simpler of the two relevant contours at cylinder locations of the \textit{inital} copies of the CFT. Next we compute the more complicated contours.

\subsubsection{Contours containing $\sum_{k=0}^{2}(t_1-t_2)^{-k-2}(-1)^k(t-t_2)(t-t_1)^k$}

Here we compute the more complicated contours at initial locations on the cylinder. For this case we again have contours of the form:

\bea
\text{Copy 1}&:&{1\over 2 \pi i}\int_{\s = 0,\t<\t_1}^{\s = 2\pi}\diff w \sum_{k=0}^{2}(t_1-t_2)^{-k-2}(-1)^k(t_+-t_2)(t_+-t_1)^k G^{(1),-}_{\dot{C}}(w)
\cr
\cr
\text{Copy 2}&:&{1\over 2 \pi i}\int_{\s = 0,\t<\t_1}^{\s = 2\pi}\diff w \sum_{k=0}^{2}(t_1-t_2)^{-k-2}(-1)^k(t_--t_2)(t_--t_1)^k G^{(2),-}_{\dot{C}}(w)
\label{Copy 1 initial}
\eea
where we again take the $t_+$ solution for Copy 1 initial and the $t_-$ solution for Copy 2 initial. For the above contours we again need to compute the integrand
\bea
&&\sum_{k=0}^{2}(-1)^k(t_1-t_2)^{-k-2}(t_{\pm}-t_2)(t_\pm-t_1)^k\cr
&&\quad\quad = {1\over 16 t_2^4}\bigg(t_{\pm} (z^2 - z A + B)-zt_2^2 + z_1t_1^2-8t_2^3\bigg)
\label{expansion of integrand two}
\eea
but now using initial state expansions. Inserting the initial state expansions of $t_{\pm}$ given in (\ref{t pm one initial}) into (\ref{expansion of integrand 2}), again with considerable simplification, we obtain:
\bea
&&\sum_{k=0}^{2}(-1)^k(t_1-t_2)^{-k-2}(t_{\pm}-t_2)(t_\pm-t_1)^k\cr
&&\quad\quad  =  {1\over 32  t_2^4}\bigg[ z^3-A'z^2+B'z + C'\cr
&&\qquad\qquad\qquad\qquad{} \mp \bigg(  \sum_{n = 0}^{\infty}C_n  z^{n + 2} -  A\sum_{n = 0}^{\infty}C_n  z^{n + 1} +B \sum_{n = 0}^{\infty}C_n  z^{n}\bigg)\bigg]\nn
\label{t pm two initial z}
\eea
Inserting $z=e^w$ into (\ref{t pm two initial z}) gives the expression:
\bea
&&\sum_{k=0}^{2}(-1)^k(t_1-t_2)^{-k-2}(t_{\pm}-t_2)(t_\pm-t_1)^k\cr
&&\quad\quad  =  {1\over 32  t_2^4}\bigg[ e^{3w} - A'e^{2w} + B'e^w + C'\cr
&&\qquad\qquad\qquad\qquad{} \mp \bigg(  \sum_{n = 0}^{\infty}C_n  e^{(n + 2)w} -  A\sum_{n = 0}^{\infty}C_n  e^{(n + 1)w} +B \sum_{n = 0}^{\infty}C_n  e^{nw}\bigg)\bigg]\nn
\label{t pm two initial}
\eea
where we have again written our expression in terms of the cylinder coordinate $w$.
We again adjust the above sums in order to combine terms. Adjusting all three of the sums in (\ref{t pm two initial}) gives:
\bea
\sum_{n = 0}^{\infty}C_{n}  e^{(n + 2)w } &=& \sum_{n = 2}^{\infty}C_{n-2}  e^{nw} 
\cr
\cr
A\sum_{n = 0}^{\infty}C_n  e^{(n + 1)w} &=& A\sum_{n = 1}^{\infty}C_{n-1}  e^{nw}\cr
&=& AC_0e^w + A\sum_{n = 2}^{\infty}C_{n-1}  e^{nw}\cr
&=& Ae^w + A\sum_{n = 2}^{\infty}C_{n-1}  e^{nw}
\cr
B\sum_{n = 0}^{\infty}C_n  e^{nw}&=&BC_0 + BC_1e^w +  B\sum_{n = 2}^{\infty}C_{n}  e^{nw }\cr
&=& B - B(a+b)e^w +  B\sum_{n = 2}^{\infty}C_{n}  e^{nw }
\label{sum initial}
\eea
Inserting the expansions in (\ref{sum initial}) into (\ref{t pm two initial}) gives:
\bea
&&\sum_{k=0}^{2}(-1)^k(t_1-t_2)^{-k-2}(t_{\pm}-t_2)(t_\pm-t_1)^k\cr
&&\quad\quad=  {1\over 32 t_2^4}\bigg[ e^{3w} - A'e^{2w} + B'e^w + C' \mp \bigg(B - (A + B(a+b))e^w \cr
&&\quad\quad\quad\quad\quad\quad\quad\quad +  \sum_{n = 2}^{\infty}\bigg(C_{n-2} - A C_{n-1} + B C_n \bigg) e^{nw} \bigg) \bigg]
\cr
&&\quad\quad = {1\over 32 t_2^4}\bigg[ e^{3w} - A'e^{2w} + B'e^w + C' \mp  \bigg( B - (A-B(a+b))e^w +  \sum_{n = 2}^{\infty}C''_n e^{nw}\bigg) \bigg]\nn
\label{t pm three initial}
\eea
where we define
\bea
C''_n\equiv C_{n-2} - A C_{n-1} + B C_n
\label{Cn prime prime}
\eea
We see that we have again writtin our expression in terms of a single infinite sum.

Inserting the expression (\ref{t pm three initial}) into (\ref{Copy 1 initial}) and using the mode definitions defined in (\ref{cylinder G initial}) we obtain the following result for Copy 1 and Copy 2 initial contours:
\bea
\text{Copy 1}&:&{1\over 2 \pi i}\int_{\s = 0,\t<\t_1}^{\s = 2\pi}\diff w \sum_{k=0}^{2}(t_1-t_2)^{-k-2}(-1)^{k}(t-t_2)(t-t_1)^k G^{(1)-}_{\dot{C}}(w)\cr
&&\qquad\qquad = {1\over  32 t_2^4}\bigg( G^{(1)-}_{ \dot{C},3} -A'G^{(1)-}_{ \dot{C},2} +B'G^{(1)-}_{ \dot{C},1} + C'G^{(1)-}_{ \dot{C},0} \cr
&&\qquad\qquad\qquad\qquad - BG^{(1)-}_{\dot{C},0} + (A + B(a+b))G^{(1)-}_{\dot{C},1} -  \sum_{n = 0}^{\infty}C''_n G^{(1)-}_{ \dot{C},n} \bigg)
\cr
\cr
\text{Copy 2}&:&{1\over 2 \pi i}\int_{\s = 0,\t<\t_1}^{\s = 2\pi}\diff w \sum_{k=0}^{2}(t_1-t_2)^{-k-2}(-1)^{k}(t-t_2)(t-t_1)^k G^{(2)-}_{\dot{C}}(w)\cr
&&\qquad\qquad = {1\over 32 t_2^4}\bigg( G^{(2)-}_{ \dot{C},3} -A'G^{(2)-}_{ \dot{C},2} +B'G^{(2)-}_{ \dot{C},1} + C'G^{(2)-}_{ \dot{C},0}\cr
&&\qquad\qquad\qquad\qquad + BG^{(2)-}_{\dot{C},0} - (A + B(a+b))G^{(2)-}_{\dot{C},1} + \sum_{n = 0}^{\infty}C''_n G^{(2)-}_{ \dot{C},n} \bigg)\nn
\label{initial contours}
\eea
Combining both contours in (\ref{initial contours}) again using the basis defined in (\ref{new basis}) gives:
\bea
&&{1\over 2 \pi i}\int_{\s = 0,\t<\t_1}^{\s = 2\pi}\diff w \sum_{k=0}^{2}(t_1-t_2)^{-k-2}(-1)^{k}(t-t_2)(t-t_1)^k G^{(1)-}_{\dot{C}}(w)\cr
&&\quad + {1\over 2 \pi i}\int_{\s = 0,\t<\t_1}^{\s = 2\pi}\diff w \sum_{k=0}^{2}(t_1-t_2)^{-k-2}(-1)^{k}(t-t_2)(t-t_1)^k G^{(2)-}_{\dot{C}}(w)
\cr
\cr
&&\quad =  {1\over  32 t_2^4}\bigg( G^{-}_{ \dot{C},3} -A'G^{-}_{ \dot{C},2} + B'G^{-}_{ \dot{C},1} + C'G^{-}_{ \dot{C},0}\cr
&&\qquad\qquad\qquad\qquad{} - B\tilde{G}^{-}_{\dot{C},0} + (A + B(a+b))\tilde{G}^{-}_{\dot{C},1} -   \sum_{n = 0}^{\infty}C''_n \tilde{G}^{-}_{ \dot{C},n} \bigg)\nn
\label{combined initial}
\eea

We note here that our expression contains a finite number of symmetric annihilation modes before the twist as well as an infinite number of antisymmetric annihilation modes before the twist. Again, since annihilation modes annihilate when acting on a state to the right we need not worry about getting an infinite number of terms.
We have now computed the two relevant types contours at locations on the cylinder corresponding to Copy 1 and Copy 2 \textit{initial} states. 

Since we have computed the relevant supercharge modes at initial and final state locations on the cylinder appearing in the expressions for our two deformation operators we are now in position to compute the final result on the cylinder.

\section{Final Result}\label{final result}
We are now able to write the expression of our two deformation operators in terms cylinder modes. Let us now rewrite (\ref{full deformation two}) in terms of $G$ modes on the cylinder which we computed in Section \ref{cylinder modes}. We recall that we have two cases to compute. The first case is when $\dot{A}=\dot{B}$ and the second case for general $\dot{A}$ and $\dot{B}$. We remind the reader that the general case reduces to the case of $\dot{A}=\dot{B}$ but case is much easier computed while in the $t$ plane. We write our two cases below.
\subsection{$\dot{A}=\dot{B}$}
Here we compute the final expression for the case where $\dot{A}=\dot{B}$. 
First let us define a shorthand notation for our twists:
\bea
\hat{\s} = \s_{2}^+(w_2)\s_{2}^+(w_1)
\label{sigma defintion}
\eea
Now inserting (\ref{t minus t2 final 2}), (\ref{t minus t2 initial 2}), and (\ref{sigma defintion}) into (\ref{A = B two}) and using (\ref{redefinition}) we obtain:
\bea
\hat{O}_{\dot{A}}\hat{O}_{\dot{A}}\!\!&=&\!\!  -{1\over 4t_2}\bigg[ G^{-}_{\dot{A},0} \bigg ( G^{-}_{\dot{A},1}   +  \sum_{n = 0}^{\infty}  C_n \tilde{G}^{-}_{\dot{A},1-n} \bigg)~\hat{\s} -  G^{-}_{\dot{A},0}~ \hat{\s}~\bigg ( G^{-}_{\dot{A},1}   -  \sum_{n = 0}^{\infty}  C_n \tilde{G}^{-}_{\dot{A},n} \bigg) 
\cr
\cr
&&\qquad{} + \bigg ( G^{-}_{\dot{A},1}  +  \sum_{n = 0}^{\infty}  C_n \tilde{G}^{-}_{\dot{A},1-n} \bigg)~\hat{\s}~ G^{-}_{\dot{A},0}  - \hat{\s}~\bigg ( G^{-}_{\dot{A},1} -  \sum_{n = 0}^{\infty}  C_n \tilde{G}^{-}_{\dot{A},n} \bigg) G^{-}_{\dot{A},0}\bigg]\nn
\eea
\subsection{General $\dot{A}$ and $\dot{B}$} 
For the case for general $\dot{A}$ and $\dot{B}$, inserting (\ref{t minus t2 final 2}), (\ref{combined final}), (\ref{t minus t2 initial 2}), (\ref{combined initial}) and (\ref{sigma defintion}) into (\ref{full deformation two}) and using the expression
\bea
C' = 2z_1t_2^2 - (a+b)B - 16 t_2^3,
\eea 
gives the final result:
\bea
\hat{O}_{\dot{B}}\hat{O}_{\dot{A}}\!\! &=&\!\! - {1\over  64 t_2^4}\bigg( G_{\dot{B},3}^{-} - A'G_{\dot{B},2}^{-} + B'G_{\dot{B},1}^{-} + \big( 2z_1t_2^2 -B(a+b)\big) G_{\dot{B},0}^{-}\cr
&&\qquad\qquad {}+  \tilde{G}_{\dot{B},3}^{-} - A'\tilde{G}_{\dot{B},2}^{-} + \sum_{n=0}^{\infty}C'_n \tilde{G}_{\dot{B},1-n}^{-}\bigg)\cr
&&\qquad\qquad\!\!\!\!\!\!\!\times ~\bigg ( G^{-}_{\dot{A},1} - e^{\Delta w\over 2}G^{-}_{\dot{A},0}  +  \sum_{n = 0}^{\infty}  C_n \tilde{G}^{-}_{\dot{A},1-n} \bigg)~\hat{\s}
\cr
\cr
\cr
&&\!\!+{1\over  64 t_2^4}\bigg( G_{\dot{B},3}^{-} - A'G_{\dot{B},2}^{-} + B'G_{\dot{B},1}^{-} + \big( 2z_1t_2^2 -B(a+b)\big)G_{\dot{B},0}^{-}\cr
&&\qquad\qquad{} +  \tilde{G}_{\dot{B},3}^{-} - A'\tilde{G}_{\dot{B},2}^{-} + \sum_{n=0}^{\infty}C'_n \tilde{G}_{\dot{B},1-n}^{-}\bigg)\cr
&&\qquad\qquad\!\!\!\!\!\!\!\times ~\hat{\s}~\bigg( G^{-}_{\dot{A},1} - e^{\Delta w\over 2}G^{-}_{\dot{A},0}  -  \sum_{n = 0}^{\infty}  C_n \tilde{G}^{-}_{\dot{A},n} \bigg)
\cr
\cr
\cr
&&\!\!- {1\over  64 t_2^4}\bigg ( G^{-}_{\dot{A},1} - e^{\Delta w\over 2}G^{-}_{\dot{A},0}  +  \sum_{n = 0}^{\infty}  C_n \tilde{G}^{-}_{\dot{A},1-n} \bigg)
~\hat{\s}
\cr
&&\qquad\qquad\!\!\!\!\!\!\!\!\!\times~\bigg( G^{-}_{ \dot{B},3} -A'G^{-}_{ \dot{B},2} + B'G^{-}_{ \dot{B},1} + \big(  2z_1t_2^2 - (a+b)B \big)G^{-}_{ \dot{B},0}\cr
&&\qquad\qquad  - B\tilde{G}^{-}_{\dot{B},0}  + (A + B(a+b))\tilde{G}^{-}_{\dot{B},1} -   \sum_{n = 2}^{\infty}C''_n \tilde{G}^{-}_{ \dot{B},n} \bigg)
\cr
\cr
\cr
&&\!\!+{1\over  64 t_2^4}~\hat{\s}~
 \bigg ( G^{-}_{\dot{A},1} - e^{\Delta w\over 2}G^{-}_{\dot{A},0}  -  \sum_{n = 0}^{\infty}  C_n \tilde{G}^{-}_{\dot{A},n} \bigg)\cr
&&\qquad\qquad\!\!\!\!\!\!\times~\bigg( G^{-}_{ \dot{B},3} -A'G^{-}_{ \dot{B},2} + B'G^{-}_{ \dot{B},1} + \big( 2z_1t_2^2 - (a+b)B \big)G^{-}_{ \dot{B},0}\cr
&&\qquad\qquad - B\tilde{G}^{-}_{\dot{B},0} + (A + B(a+b))\tilde{G}^{-}_{\dot{B},1}  -   \sum_{n = 2}^{\infty}C''_n \tilde{G}^{-}_{ \dot{B},n} \bigg)\nn
\label{final result one}
\eea
Writing our final result in compact notation, (\ref{final result one}) becomes:
\bea
\hat{O}_{\dot{B}}\hat{O}_{\dot{A}} &=& -{1\over 64 t_2^4}\bigg[\bigg(\sum_{n=0}^{3} D_{n}G^{-}_{\dot{B},n} + \sum_{n=-2}^{\infty}E_{n}\tilde{G}^{-}_{\dot{B},1-n} \bigg)\bigg(\sum_{n=0}^1 D'_n G^-_{\dot{A},n} + \sum_{n=0}^{\infty} C_n\tilde{G}^-_{\dot{A},1-n}\bigg)~\hat{\s}\cr
&&\qquad\quad -\bigg(\sum_{n=0}^{3} D_{n}G^{-}_{\dot{B},n} + \sum_{n=-2}^{\infty}E_{n}\tilde{G}^{-}_{\dot{B},1-n} \bigg)~\hat{\s}~\bigg(\sum_{n=0}^1 D'_n G^-_{\dot{A},n} - \sum_{n=0}^{\infty} C_n\tilde{G}^-_{\dot{A},n}\bigg)\cr
&&\qquad\quad + \bigg(\sum_{n=0}^1 D'_n G^-_{\dot{A},n} + \sum_{n=0}^{\infty} C_n\tilde{G}^-_{\dot{A},1-n}\bigg)~\hat{\s}~ \bigg(\sum_{n=0}^{3} D_{n}G^{-}_{\dot{B},n} + \sum_{n=0}^{\infty}E'_{n}\tilde{G}^{-}_{\dot{B},n} \bigg)\cr
&&\qquad\quad -\hat{\s}~\bigg(\sum_{n=0}^1 D'_n G^-_{\dot{A},n} - \sum_{n=0}^{\infty} C_n\tilde{G}^-_{\dot{A},n}\bigg) \bigg(\sum_{n=0}^{3} D_{n}G^{-}_{\dot{B},n} + \sum_{n=0}^{\infty}E'_{n}\tilde{G}^{-}_{\dot{B},n} \bigg)\bigg]\nn
\eea
where we have computed the above coefficients in terms of $\Delta w$ in detail in Appendix \ref{coefficients}. We tabulate the results below:
\bea
C_n &=& (-1)^{n}\sum_{k = - {n\over 2}}^{{n \over 2}}{}^{1/2}C_{{n\over 2} + k}{}^{1/2}C_{{n\over 2} - k}\cosh\big( k \Delta w\big)\cr
D_0 &=& -{1\over 4}e^{-{3\Delta w\over2}}\big( 1 + 3e^{\Delta w} \big)\cr
D_1 &=&{3\over 2}\big( 1 + e^{-\Delta w} \big)\cr
D_2 &=& -{3\over 4}e^{-{\Delta w\over 2}}\big( 3 + e^{\Delta w} \big)\cr
D_3 &=&  1\cr
\cr
D'_0 &=& -e^{\Delta w \over 2}\cr
D'_1 &=& 1\cr
\cr
E_{-2}&=& 1\cr
E_{-1}&=& -{3\over 4}e^{-{\Delta w\over 2}}\big( 3 + e^{\Delta w} \big)\cr
E_{n} &=&  \bigg(C_{n+2}- {1\over 4}e^{-{\Delta w\over 2}}\big( 7 + e^{\Delta w} \big) C_{n+1} +{1\over 4}\big(1 + 3e^{-\Delta w} \big) C_n\bigg),\quad n\geq 0\cr
\cr
E'_0 &=& -{1\over 4}\big(1 + 3e^{-\Delta w} \big)\cr
E'_1 &=& {3\over4} e^{-{\Delta w\over 2}} \big(3 + \cosh\big(\Delta w\big)\big)\cr
E'_n &=& -\bigg( C_{n-2} - {1\over 4}e^{-{\Delta w\over 2}}\big( 7 + e^{\Delta w} \big) C_{n-1} + {1\over 4}\big(1 + 3e^{-\Delta w} \big)C_n\bigg),\quad n\geq 2
\eea
We see that in both cases above, we were able to write the full expression of two deformation operators in terms of modes before and after the twists using the covering space method.

Also, we again note that there are a finite number of symmetric and antisymmetric annihilation modes after the twist and a finite number of symmetric annihilation modes before the twist but an infinite number of antisymmetric creation modes after the twist and an infinite number of antisymmetric annihilation modes before the twist. These terms all give finite contributions when computing an amplitude because creators kill on the left and annihilators kill on the right. We need not worry about an infinite number of contributions when capping each side with an appropriate state.

\subsection{Numerical Plots of $C_{n}$}
Here we numerically plot the coefficient
\bea
C_n = \sum_{k=-{n\over2}}^{n\over2}{}^{\h}C_{{n\over2}+k}{}^{\h}C_{{n\over2}-k}\cosh(k\D w)
\label{C euclidean}
\eea 
to determine its behavior for various mode numbers, $n$.
We can write our twist separation $\Delta w$ as:
\bea
\Delta w = \Delta \t + i\Delta \s
\eea
We wick rotate back to Minkowski signature by taking $\Delta \t\to i\Delta t$ which gives
\bea
\Delta w \to i \Delta w
\label{dw}
\eea
with $\Delta w = \Delta t + \Delta \s$ being completely real. Inserting (\ref{dw}) into (\ref{C euclidean}) gives
\bea
C_n = \sum_{k=-{n\over2}}^{n\over2}{}^{\h}C_{{n\over2}+k}{}^{\h}C_{{n\over2}-k}\cos(k\D w)
\label{C minkowksi}
\eea 
We now plot (\ref{C minkowksi}) in Figure \ref{figtwo} for various values of $n$:
\begin{figure}[tbh]
\begin{center}
\includegraphics[width=0.42\columnwidth]{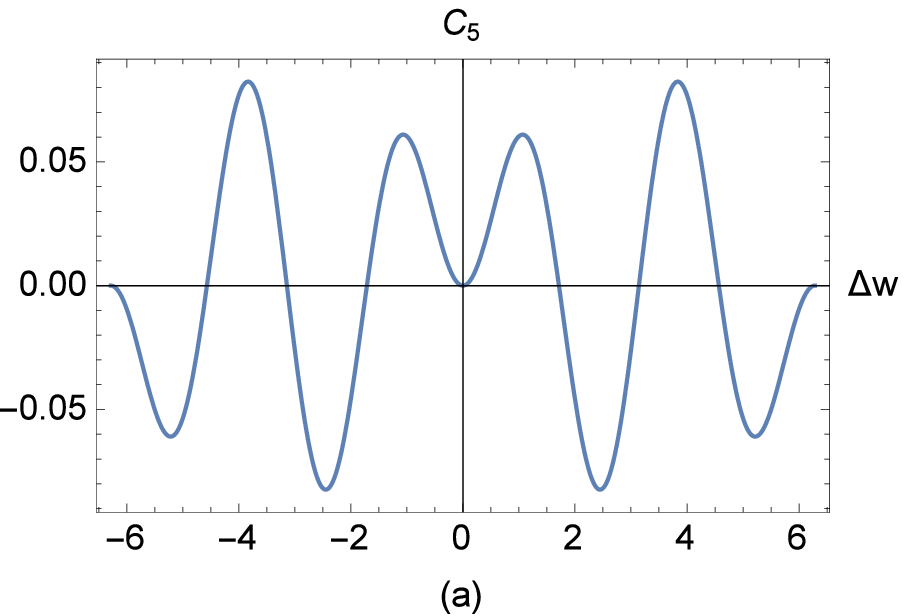} $\qquad\qquad$ \includegraphics[width=0.42\columnwidth]{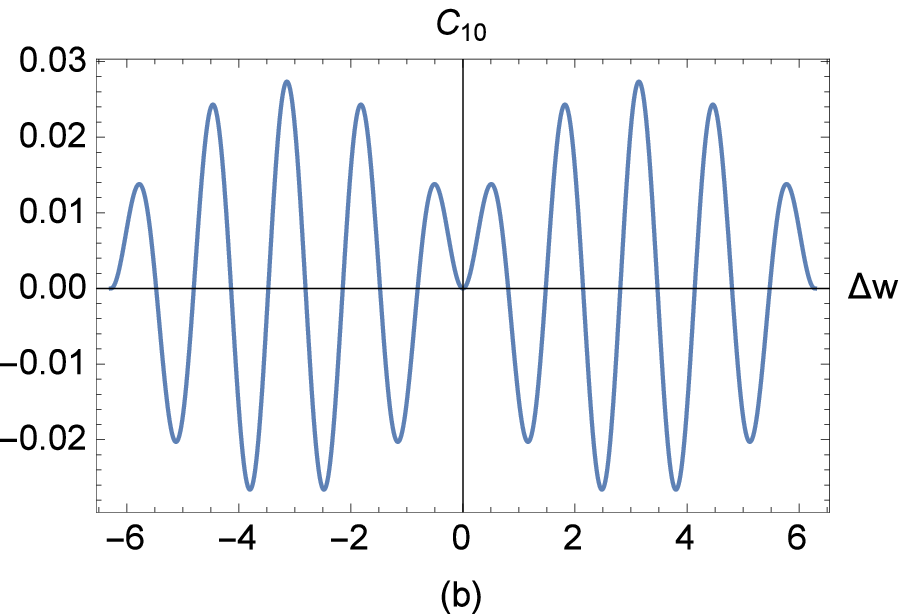} $\qquad\qquad$
\includegraphics[width=0.42\columnwidth]{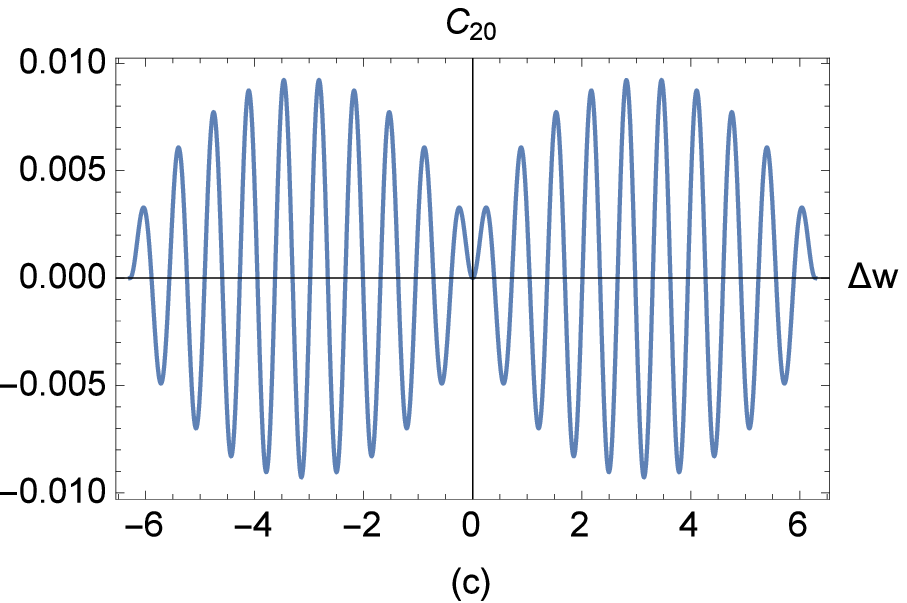}
\end{center}
\caption{We plot $C_n$ for (a) $n=5$ (b) $n=10$ and (c) $n=20$ each from $\Delta w \in [-2\pi,2\pi]$}
\label{figtwo}
\end{figure}

We see an oscillatory behavior as well as an envelope behavior coming from the twist separation, $\Delta w$. The number of peaks from $0$ to $2\pi$ is given by ${ n\over 2}$ for even $n$ and ${n-1\over 2}$ for odd $n$.

\section{Discussion}

Black hole formation is a process that has yet to possess a full quantitative description. Because this process is difficult to study in the gravity setting, it is useful to try to gain some insight using the dual CFT.  Generally, the process of black hole formation in the gravity theory would manifest as the process of thermalization in the CFT. Looking at the D1D5 CFT, we investigate this question of thermalization by considering a deformation away from its conjectured `orbifold point' where the theory is free.\footnote{For another approach to thermalization in the CFT, see \cite{kaplan}.} This deformation, $\hat{O}_{\dot{A}}$, consists of a supercharge $G$ and a twist $\s$. Looking at first order in the deformation operator, one finds no clear evidence of thermalization. We thus extended our analysis to second order in the twist operator where the first twist joins two singly wound copies into one doubly wound copy and then the second twist returns the double wound copy back to two singly wounds copies. One finds  that after applying two twists to the vacuum a squeezed state is produced just as in the one twist case. In prior work, we computed the bosonic and fermionic Bogoluibov coefficients $\g^{B},\g^{F\pm}$ describing this squeezed state. In the large $m,n$ limit we found a decay with $m,n$ similar to the one twist case \cite{chmt1,cmt}, but also found  an additional oscillatory dependence arising from  the twist separation parameter $\Delta w$ \cite{chm1}. We conjectured that to all orders in the deformation, the Bogoliubov coefficients could be factored into  a part independent of the twist separations, and a part that oscillated with these separations.\footnote{For other computations with the twist deformation in the CFT, see \cite{peet}.}

In this work we extended our analysis further by computing the action of the supercharge, $G$, contained in the full deformation, $O_{\dot{B}}O_{\dot{A}}$. This involved stretching the $G$ contours off of the twists, $\s_2$, to initial and final states on the cylinder. We began the computation on the cylinder where we first stretched the second $G$ contour off of the second twist producing three separate terms. In the first term the second $G$ contour was stretched above both twists with the first $G$ contour wrapping the first twist. In the second term both $G$ contours wrapped the first twist. In the third term the second $G$ contour stretched below both twists with the first $G$ contour wrapping the first twist. However, the goal was to write the full deformation operator completely in terms of contours before and after both twists with nothing in between. In order to do this we mapped these three terms into the covering $t$ plane. We then stretched the $G$ contours in each of the three terms to initial state and final state punctures in the t plane representing initial and final states on cylinder. We then mapped this result back to the cylinder to obtain the final expression for both deformation operators. We note that the final expression is quite complicated and lengthy. This is because there are many possible $t$ plane locations for the $G$ contours to land on.  We found that the modes of the supercharge involved functions like $C_n(\Delta w)$ which had the form of a rapidly oscillating function inside a smooth envelope. 

Now that we have the complete action of the deformation operator at second order, we hope to be able to proceed to the next steps required to find thermalization in the CFT. Our expression can be applied to any initial state.  One can choose a simple state consisting of a single left and a single right oscillator, which can represent one high energy particle thrown into the AdS throat. One must then compute the amplitude that we have computed in the present paper for an arbitrary separation between the two twists, and integrate over the locations of these twists. The result would represent the breakup of the initial excitation into lower energy excitations. This is the essential vertex describing thermalization, in the sense that each of the resulting oscillators can be further broken up into lower energy excitations in the same way, and so on. Thus multiple applications of this vertex gives the decay of a high energy particle into the low energy has of excitations that would describe the black hole phase. Of course there will be more complicated effects coming from the interaction of three twists and so on, but we expect that the qualitative picture of thermalization would emerge by considering the repeated application of the two-twist vertex. We hope to return in a following paper to complete these remaining steps.

\section*{Acknowledgements}
This work is supported in part by DOE grant de-sc0011726.\\\\

\appendix
\section{CFT notation and conventions} \label{ap:CFT-notation}

We follow the notation of \cite{acm1, acm2}, which we record here for convenience.
We have 4 real left moving fermions $\psi_1, \psi_2, \psi_3, \psi_4$ which we group into doublets $\psi^{\alpha A}$ as follows:
\be
\begin{pmatrix}
\psi^{++} \cr \psi^{-+}
\end{pmatrix}
=\sqi
\begin{pmatrix}
\psi_1+i\psi_2 \cr \psi_3+i\psi_4
\end{pmatrix}
\ee
\be
\begin{pmatrix}
\psi^{+-} \cr \psi^{--}
\end{pmatrix}
=\sqi
\begin{pmatrix}
\psi_3-i\psi_4 \cr -(\psi_1-i\psi_2)
\end{pmatrix}.
\ee
Here $\alpha=(+,-)$ is an index of the subgroup $SU(2)_L$ of rotations on $S^3$ and $A=(+,-)$ is an index of the subgroup $SU(2)_1$ from rotations in $T^4$. The reality conditions on the individual fermions are
\bea
(\psi_i)^{\dagger} = \psi_i \qquad \Rightarrow \qquad (\psi^{\a A})^{\dagger} = - \epsilon_{\alpha\beta}\epsilon_{AB} \psi^{\beta B} \,.
\eea 
One can introduce doublets $\psi^\dagger$, whose components are given by
\bea
(\psi^\dagger)_{\alpha A} &=& (\psi^{\a A})^{\dagger},
\eea 
from which the reality condition is given by
\be
 (\psi^\dagger)_{\alpha A}=-\epsilon_{\alpha\beta}\epsilon_{AB} \psi^{\beta B}.
\ee
The 2-point functions are
\be
<\psi^{\alpha A}(z)(\psi^\dagger)_{\beta B}(w)>=\delta^\alpha_\beta\delta^A_B{1\over z-w}, ~~~
<\psi^{\alpha A}(z)\psi^{\beta B}(w)>=-\epsilon^{\alpha\beta}\epsilon^{AB}{1\over z-w},
\ee
where we have:
\be
\epsilon_{12}=1, ~~~\epsilon^{12}=-1, ~~~
\psi_A=\epsilon_{AB}\psi^B, ~~~
\psi^A=\epsilon^{AB}\psi_B \,.
\ee
There are 4 real left moving bosons $X_1, X_2, X_3, X_4$, which can be grouped into a matrix:
\be
X_{A\dot A}= \sqi X_i \sigma_i
=\sqi
\begin{pmatrix}
X_3+iX_4 & X_1-iX_2 \\ X_1+iX_2&-X_3+iX_4
\end{pmatrix},
\ee
where $\sigma_i=(\sigma_a, iI)$. The reality condition on the individual bosons is given by
\bea
(X_i)^{\dagger} = X_i \qquad \Rightarrow \qquad (X_{A\dot A})^{\dagger} = - \e^{AB}\e^{\dot A \dot B} X_{B \dot B} \,.
\eea 
One can introduce a matrix, $X^\dagger$, with components  
\be
(X^\dagger)^{A\dot A}~=~ (X_{A\dot A})^{\dagger}~=~\sqi
\begin{pmatrix}
X_3-iX_4& X_1+iX_2\\
X_1-iX_2&-X_3-iX_4
\end{pmatrix},
\ee
from which the reality condition is given by
\bea
(X^\dagger)^{A\dot A}~=~ - \e^{AB}\e^{\dot A \dot B} X_{B \dot B} \,.
\eea 
The 2-point functions are
\be
<\partial X_{A\dot A}(z) (\partial X^\dagger)^{B\dot B}(w)>=-{1\over (z-w)^2}\delta^B_A\delta^{\dot B}_{\dot A}, ~~~
<\partial X_{A\dot A}(z) \partial X_{B\dot B}(w)>={1\over (z-w)^2}\epsilon_{AB}\epsilon_{\dot A\dot B} \,.
\ee

The chiral algebra is generated by the operators
\be
J^a=-{1\over 4}(\psi^\dagger)_{\alpha A} (\sigma^{Ta})^\alpha{}_\beta \psi^{\beta A}
\ee
\be
G^\alpha_{\dot A}= \psi^{\alpha A} \pa X_{A\dot A}, ~~~(G^\dagger)_{\alpha}^{\dot A}=(\psi^\dagger)_{\alpha A} \pa (X^\dagger)^{A\dot A}
\ee
\be
T=-{1\over 2} (\pa X^\dagger)^{A\dot A}\pa X_{A\dot A}-{1\over 2} (\psi^\dagger)_{\alpha A} \pa \psi^{\alpha A}
\ee
\be
(G^\dagger)_{\alpha}^{\dot A}=-\epsilon_{\alpha\beta} \epsilon^{\dot A\dot B}G^\beta_{\dot B}, ~~~~G^{\alpha}_{\dot A}=-\epsilon^{\alpha\beta} \epsilon_{\dot A\dot B}(G^\dagger)_\beta^{\dot B} \,.
\ee
These operators generate the OPE algebra
\be
J^a(z) J^b(z')\sim \delta^{ab} {\h\over (z-z')^2}+i\epsilon^{abc} {J^c\over z-z'}
\ee
\be
J^a(z) G^\alpha_{\dot A} (z')\sim {1\over (z-z')}\h (\sigma^{aT})^\alpha{}_\beta G^\beta_{\dot A}
\ee
\be
G^\alpha_{\dot A}(z) (G^\dagger)^{\dot B}_\beta(z')\sim -{2\over (z-z')^3}\delta^\alpha_\beta \delta^{\dot B}_{\dot A}- \delta^{\dot B}_{\dot A}  (\sigma^{Ta})^\alpha{}_\beta [{2J^a\over (z-z')^2}+{\p J^a\over (z-z')}]
-{1\over (z-z')}\delta^\alpha_\beta \delta^{\dot B}_{\dot A}T
\ee
\be
T(z)T(z')\sim {3\over (z-z')^4}+{2T\over (z-z')^2}+{\p T\over (z-z')}
\ee
\be
T(z) J^a(z')\sim {J^a\over (z-z')^2}+{\p J^a\over (z-z')} 
\ee
\be
T(z) G^\alpha_{\dot A}(z')\sim {{3\over 2}G^\alpha_{\dot A}\over (z-z')^2}  + {\p G^\alpha_{\dot A}\over (z-z')} \,.
\ee

Note that
\be
J^a(z) \psi^{\gamma C}(z')\sim {1\over 2} {1\over z-z'} (\sigma^{aT})^\gamma{}_\beta \psi^{\beta C} \,.
\ee

The above OPE algebra gives the commutation relations
\begin{eqnarray}
\com{J^a_m}{J^b_n} &=& \frac{m}{2}\delta^{ab}\delta_{m+n,0} + i{\epsilon^{ab}}_c J^c_{m+n}
            \\
\com{J^a_m}{G^\alpha_{\dot{A},n}} &=& \frac{1}{2}{(\sigma^{aT})^\alpha}_\beta G^\beta_{\dot{A},m+n}
             \\
\ac{G^\alpha_{\dot{A},m}}{G^\beta_{\dot{B},n}} &=& \hspace*{-4pt}\epsilon_{\dot{A}\dot{B}}\bigg[
   (m^2 - \frac{1}{4})\epsilon^{\alpha\beta}\delta_{m+n,0}
  + (m-n){(\sigma^{aT})^\alpha}_\gamma\epsilon^{\gamma\beta}J^a_{m+n}
  + \epsilon^{\alpha\beta} L_{m+n}\bigg]\quad\\
\com{L_m}{L_n} &=& \frac{m(m^2-\frac{1}{4})}{2}\delta_{m+n,0} + (m-n)L_{m+n}\\
\com{L_m}{J^a_n} &=& -n J^a_{m+n}\\
\com{L_m}{G^\alpha_{\dot{A},n}} &=& \left(\frac{m}{2}-n\right)G^\alpha_{\dot{A},m+n} \,.
\end{eqnarray}

\section{Rewriting the Final Result in terms of $\Delta w$}\label{coefficients}
Here we compute the expressions for all of the various coefficients in our final expression for the full second order deformation operator, $\hat{O}_{\dot{B}}\hat{O}_{\dot{A}}$, in terms of $\Delta w$ which is the only free parameter. First let us remind the reader of the following parameters
\bea
z_1&=&e^{-{\Delta w\over 2}}\cr
z_2&=&e^{\Delta w\over 2} \cr
a&=&\cosh^2\bigg({\Delta w\over 4}\bigg)\cr
b&=&\sinh^2\bigg({\Delta w\over 4}\bigg)\cr
t_2&=&\sqrt{ab}\cr
&=&{1\over 2}\sinh\bigg({\Delta w\over 2}\bigg)
\label{coefficients one}
\eea
Now let us write our final expression below:
\bea
\hat{O}_{\dot{B}}\hat{O}_{\dot{A}} &=& -{1\over 64 t_2^4}\bigg[\bigg(\sum_{n=0}^{3} D_{n}G^{-}_{\dot{B},n} + \sum_{n=-2}^{\infty}E_{n}\tilde{G}^{-}_{\dot{B},1-n} \bigg)\bigg(\sum_{n=0}^1 D'_n G^-_{\dot{A},n} + \sum_{n=0}^{\infty} C_n\tilde{G}^-_{\dot{A},1-n}\bigg)~\hat{\s}\cr
&&\qquad\quad -\bigg(\sum_{n=0}^{3} D_{n}G^{-}_{\dot{B},n} + \sum_{n=-2}^{\infty}E_{n}\tilde{G}^{-}_{\dot{B},1-n} \bigg)~\hat{\s}~\bigg(\sum_{n=0}^1 D'_n G^-_{\dot{A},n} - \sum_{n=0}^{\infty} C_n\tilde{G}^-_{\dot{A},n}\bigg)\cr
&&\qquad\quad + \bigg(\sum_{n=0}^1 D'_n G^-_{\dot{A},n} + \sum_{n=0}^{\infty} C_n\tilde{G}^-_{\dot{A},1-n}\bigg)~\hat{\s}~ \bigg(\sum_{n=0}^{3} D_{n}G^{-}_{\dot{B},n} + \sum_{n=0}^{\infty}E'_{n}\tilde{G}^{-}_{\dot{B},n} \bigg)\cr
&&\qquad\quad -\hat{\s}~\bigg(\sum_{n=0}^1 D'_n G^-_{\dot{A},n} - \sum_{n=0}^{\infty} C_n\tilde{G}^-_{\dot{A},n}\bigg) \bigg(\sum_{n=0}^{3} D_{n}G^{-}_{\dot{B},n} + \sum_{n=0}^{\infty}E'_{n}\tilde{G}^{-}_{\dot{B},n} \bigg)\bigg]\nn
\eea
With the following coefficient definitions:
\bea
D_0 &\equiv& \big( 2z_1t_2^2 -B(a+b)\big)\cr
D_1 &\equiv& B'\cr
D_2 &\equiv& -A'\cr
D_3&\equiv&  1\cr
\cr
D'_0 &\equiv& -e^{\Delta w \over 2}\cr
D'_1 &\equiv& 1\cr
\cr
E_{-2}&\equiv& 1\cr
E_{-1}&\equiv& -A'\cr
E_{n} &\equiv& C'_n,\quad n\geq 0\cr
\cr
E'_0 &\equiv& -B\cr
E'_1 &\equiv& A+ B(a+b)\cr
E'_n &\equiv& - C''_n,\quad n\geq 2
\label{coefficients four}
\eea
and 
\bea
C_n &\equiv& (-1)^{n}\sum_{k = - {n\over 2}}^{{n \over 2}}{}^{1/2}C_{{n\over 2} + k}{}^{1/2}C_{{n\over 2} - k}\cosh\big( k \Delta w\big)\cr
C'_n &\equiv&  (C_{n+2}- A C_{n+1} + B C_n\big)\cr
C''_n&\equiv&C_{n-2} - A C_{n-1} + B C_n
\label{coefficients C}
\eea
Where 
\bea
A &\equiv&(2z_1 + t_2)\cr
B &\equiv& z_1 (a+b)+ t_2(z_2 - 2(a+b))
\cr
A'&\equiv&A+a+b\cr
B'&\equiv&A(a+b) + B - 2t_2^2
\label{coefficients two}
\eea

Now evaluating (\ref{coefficients two}) in terms of $\Delta w$ using (\ref{coefficients one}) gives
\bea
A &=&{1\over 4}e^{-{\Delta w\over 2}}\big( 7 + e^{\Delta w} \big)\cr
\cr
B &=&{1\over 4}\big(1 + 3e^{-\Delta w} \big)
\cr
\cr
A'&=& {3\over 4}e^{-{\Delta w\over 2}}\big( 3 + e^{\Delta w} \big)\cr
\cr
B' &=& {3\over 2}\big( 1 + e^{-\Delta w} \big)
\label{coefficients three}
\eea
Inserting (\ref{coefficients three}) into (\ref{coefficients C}) where appropriate gives:
\bea
C_n &=& (-1)^{n}\sum_{k = - {n\over 2}}^{{n \over 2}}{}^{1/2}C_{{n\over 2} + k}{}^{1/2}C_{{n\over 2} - k}\cosh\big( k \Delta w\big)\cr
C'_n &=&  \bigg(C_{n+2}- {1\over 4}e^{-{\Delta w\over 2}}\big( 7 + e^{\Delta w} \big) C_{n+1} +  {1\over 4}\big(1 + 3e^{-\Delta w} \big) C_n\bigg)\cr
C''_n&=&\bigg( C_{n-2} - {1\over 4}e^{-{\Delta w\over 2}}\big( 7 + e^{\Delta w} \big) C_{n-1} + {1\over 4}\big(1 + 3e^{-\Delta w} \big)C_n\bigg)
\label{coefficients C two}
\eea
Now inserting (\ref{coefficients one}), (\ref{coefficients three}), and (\ref{coefficients C two}) into (\ref{coefficients four}) where appropriate gives:
\bea
D_0 &=& -{1\over 4}e^{-{3\Delta w\over2}}\big( 1 + 3e^{\Delta w} \big)\cr
D_1 &=&{3\over 2}\big( 1 + e^{-\Delta w} \big)\cr
D_2 &=& -{3\over 4}e^{-{\Delta w\over 2}}\big( 3 + e^{\Delta w} \big)\cr
D_3 &=&  1\cr
\cr
D'_0 &=& -e^{\Delta w \over 2}\cr
D'_1 &=& 1\cr
\cr
E_{-2}&=& 1\cr
E_{-1}&=& -{3\over 4}e^{-{\Delta w\over 2}}\big( 3 + e^{\Delta w} \big)\cr
E_{n} &=&  \bigg(C_{n+2}- {1\over 4}e^{-{\Delta w\over 2}}\big( 7 + e^{\Delta w} \big) C_{n+1} +  {1\over 4}\big(1 + 3e^{-\Delta w} \big) C_n\bigg),\quad n\geq 0\cr
\cr
E'_0 &=& -{1\over 4}\big(1 + 3e^{-\Delta w} \big)\cr
E'_1 &=& {3\over4} e^{-{\Delta w\over 2}} \big(3 + \cosh\big(\Delta w\big)\big)\cr
E'_n &=& -\bigg( C_{n-2} - {1\over 4}e^{-{\Delta w\over 2}}\big( 7 + e^{\Delta w} \big) C_{n-1} + {1\over 4}\big(1 + 3e^{-\Delta w} \big)C_n\bigg),\quad n\geq 2
\label{coefficients five}
\eea

\end{document}